\documentclass[referee]{aaph} 
\usepackage{graphicx}
\usepackage{txfonts}
\usepackage{longtable,lscape}
\usepackage{hyperref}
\usepackage{natbib}
\newcommand{\nc}{\newcommand}
\nc{\teff}{$T_{\rm{eff}}$\,}
\nc{\logg}{$\log \,(g)$\,}
\nc{\kms}{\,${\rm km\,s}^{-1}$\,}
\nc{\vsini}{$v \sin \,(i)$\,}      
\nc{\vmic}{$v_{\rm{mic}}$\,}
\nc{\vrad}{$v_{\rm{rad}}$\,}
\nc{\fe}{$\rm{[Fe/H]}$\,} 
\nc{\li}{$A(\rm{Li})$\,} 
\nc{\cen}{$\rm{LRc}01$} 
\nc{\anti}{$\rm{LRa}01$}
\nc{\uvesstars}{UVES stars }
\nc{\hydrastars}{Hydra stars }

\nc{\ALi}{$A(\rm{Li})$}
 
\begin{document}

   \title{Stellar parameters  for  stars of the CoRoT exoplanet field
    \thanks{Based on observations
    obtained with the UVES (VLT/UT2 ESO program 077.D-0446A) and Hydra/Blanco 4m (CTIO-NOAO program P\#9005) spectrographs.}}

   \author{C. Cort\'es\inst{1,2,3} 
          \and
          S. C. Maciel\inst{2}
          \and
          S. Vieira\inst{2,5}
          \and
          C. E. Ferreira Lopes\inst{2,6}
          \and 
          \\ I.~C. Le\~ao\inst{2}
          \and
          G. P. de Oliveira\inst{2}
          \and
          C. Correia\inst{2}
          \and
          B. L. Canto Martins\inst{2}
         \and
         M. Catelan\inst{3,4}    
         \and 
          J. R. De Medeiros\inst{2}  
          }
  
   \institute{Departamento de F\'{i}sica, Facultad de Ciencias B\'asicas, Universidad Metropolitana de la Educaci\'on, Av. Jos\'e Pedro Alessandri 774,7760197, Nu\~noa, Santiago, Chile \\
              \email{cristian.cortes@umce.cl} 
     \and
     Departamento de F\'{i}sica, Universidade Federal do Rio Grande do Norte, 59072-970 Natal, RN, Brazil
     \and
     Millennium Institute of Astrophysics (MAS), Santiago, Chile
     \and 
        Ponticia Universidad Cat\'olica de Chile, Instituto de Astrof\'{i}sica, Av. Vicu\~na Mackenna 4860, 782-0436 Macul, Santiago, Chile
    \and
     Universidade Federal de Roraima, Campus Paricarana, Av. Cap. Ene Garcez 2413, Aeroporto, 69304-000 Boa Vista, RR, Brazil
     \and
     Scottish Universities Physics Alliance, Wide-Field Astronomy Unit, Institute for Astronomy, School of Physics and Astronomy, University of Edinburgh, Royal Observatory, Blackford Hill, Edinburgh EH9 3HJ, UK }

   \date{accepted May 2015}
 
 
  \abstract
   {Spectroscopic observations represent a fundamental step in the physical characterization of stars and, in particular, in the
   precise location of stars in the HR diagram. Rotation is also a key parameter, impacting   stellar properties 
   and evolution, which modulates the interior and manifests itself on the surface of stars. To date, the lack of 
   analysis based on large samples has prevented our understanding of the real impact of stellar parameters and rotation 
   on the stellar evolution as well as on the behavior of surface abundances. The space missions, CoRoT and Kepler, are 
   providing us with rotation periods for thousands of stars, thus enabling a robust assessment of 
   the behavior of rotation for different populations and evolutionary stages. For these reasons, the follow-up programs 
   are fundamental to increasing the returns of these space missions. An analysis that combines spectroscopic data and 
   rotation/modulation periods obtained from these space missions  provides the basis for establishing the 
   evolutionary behavior of the angular momentum of solar-like stars at different evolutionary stages, and the relation 
   of rotation with other relevant physical and chemical parameters.
   }
   {To support the computation and evolutionary interpretation of periods associated with 
   the rotational modulation, oscillations, and variability  of stars located in the CoRoT 
   fields, we are conducting a spectroscopic survey for stars located in the fields already observed 
   by the satellite. These observations allow us to compute physical and chemical parameters for our stellar sample.
   }
   {Using spectroscopic observations obtained with UVES/VLT and Hydra/Blanco, and based on
    standard analysis techniques, we  computed physical and chemical parameters (\teff, \logg, \fe, \vmic,  \vrad, \vsini, and \li) for a large sample of CoRoT targets.
    }
   {We provide physical and chemical parameters for a sample comprised of  138 CoRoT targets. 
   Our analysis shows the stars in our sample are located in different evolutionary stages, ranging from the main sequence to 
        the red giant branch, and range in spectral type from F to K.  The physical and chemical properties 
 for the stellar sample are in agreement with typical values reported for FGK stars. However, we report three stars presenting abnormal lithium
 behavior in the CoRoT fields.  These parameters allow us to properly characterize the intrinsic properties of the stars in these fields. 
        Our results reveal important differences in the distributions of metallicity, $T_{\rm eff}$, and evolutionary status for  stars belonging to  different 
        CoRoT fields, in agreement with results obtained independently from ground-based photometric surveys. 
        }
   {Our spectroscopic catalog, by providing much-needed spectroscopic information for a large sample of CoRoT targets, will be of 
   key importance for the successful accomplishment     of several different programs related to the CoRoT mission, thus it will help further boost
        the scientific return associated with this space mission.
        }


   \maketitle
%

\section{Introduction}

The CoRoT (Convection, Rotation, and planetary Transits) space mission \citep{Baglin_etal2007} 
  collected a total of 161,303 point-source photometric data over a period of six years for stars exhibiting different 
luminosity classes and spectral types.
This space mission  had two main goals: 1)~the detection 
of extra-solar planets using the transit procedure, and 2)~precise stellar seismology. In addition to 
these two big challenges, other programs related to the CoRoT mission are helping further our 
understanding of a variety of important astrophysical phenomena, such as stellar activity, pulsation, 
and multiplicity. In this context, CoRoT provides a unique opportunity for the study of stellar 
rotation,  which is recognized as a fundamental quantity that controls the 
evolution of stars, however, it is only today that the models and observations in hand to begin to address it. 
In this sense, CoRoT offers the necessary tools for the photometric measurements 
of rotation periods for a statistically robust sample of stars at different evolutionary stages 
and belonging to different stellar populations. 

However, in spite of the large amount of high-quality photometric data obtained with CoRoT, ground-based 
observations are also needed to accomplish the main scientific goals of the mission. For this reason, 
several observational programs have increased the photometric database related to the CoRoT mission 
\citep[e.g.,][]{Aigrain_etal2009,Deleuil_etal2009}. These data enabled the establishment of the best fields and 
observation setups to observe with the satellite,  also establishing the spectral types and 
luminosity classes for the stars in the CoRoT fields. Nevertheless, important uncertainties are still 
present in these classifications because of the variable reddening levels affecting the CoRoT fields, 
in addition to the still unknown chemical abundances and distances to the targets.       

 Spectroscopic observations are mandatory for a solid treatment of the different CoRoT 
scientific goals. In this sense, two large spectroscopic surveys of CoRoT targets have been carried out 
to date, both using multifiber observations. The first survey \citep[G10 hereafter]{Gazzano_etal2010} combined 
multifiber observations with an automated procedure for the determinations of different stellar parameters, 
whereas the second  was dedicated essentially to spectral classification \citep[]{Sebastian_etal2012}.

In the context of the physical characterization of CoRoT targets, we  carried out a large spectroscopic 
survey focused on the brightest F-, G-, and K-type stars in the CoRoT exoplanetary fields LRc01 and LRa01, 
using the multifiber spectrographs UVES/VLT and Hydra/Blanco, with high and medium spectral resolution, 
respectively. Using these observations, we applied a homogeneous procedure for the determination of 
different stellar parameters, including effective temperature (\teff), surface gravity (\logg), overall 
metallicity (\fe), radial velocity (\vrad), projected rotational velocity (\vsini), and  microturbulence 
(\vmic).  The main goal of this paper is to present the corresponding catalog. 
 We also present the mean values for stellar parameters of the two stellar populations in the CoRoT anticenter/center direction.
The paper is structured as follows: in Sect.~\ref{CAP:OBS} we describe the observations and data reduction. Sect.~3 describes how we derived the stellar parameters  for the stars in our sample, 
and Sect.~4 contains our main results. Finally, we draw our conclusions in Sect.~5.


\section{Observations}\label{CAP:OBS}

The present stellar sample is composed of  138 stars of spectral types F, G, and K, with visual 
magnitudes  $V$  between $10$ to $14$, located in two  exoplanet fields observed by CoRoT, namely the 
Galactic center (\cen: \emph{Long Run Center 01}) and the Galactic anticenter (\anti: \emph{Long 
Run Anticenter 01}) fields.   We selected the sample  
using as criteria the visual magnitude V, the spectral type, and the luminosity classes defined by \citet{Deleuil_etal2009} for CoRoT targets.
We selected stars belonging to luminosity classes II, III, IV, and V  considering the range in V and spectral type defined above. 
Our sample is comprised of the brightest stars in  both CoRoT fields, and is thus not fully representative of the 
magnitude and color distribution of CoRoT stars.

To obtain a physical characterization for these stars,  a series of spectroscopic observations 
were carried out using two spectrographs. A sample of $ 53$ stars was observed using the high-resolution 
UVES spectrograph (hereafter \uvesstars) mounted on the Kuyen/VLT 8.2m telescope, located in Cerro 
Paranal, Chile, in the course of different observing runs in 2006. The UVES standard setup DICH-2 
 ($390-760$nm)  with a $0.9$~arcsec slit was used, allowing us to obtain high-resolution ($R\approx 47,\!000$) 
and high signal-to-noise ($S/N>100$) spectra.The main characteristics of the targets and the observation 
dates are given in Table~\ref{TAB:FUVES}. 

\begin{figure}[!h]
   \centering
   \includegraphics[scale=0.6]{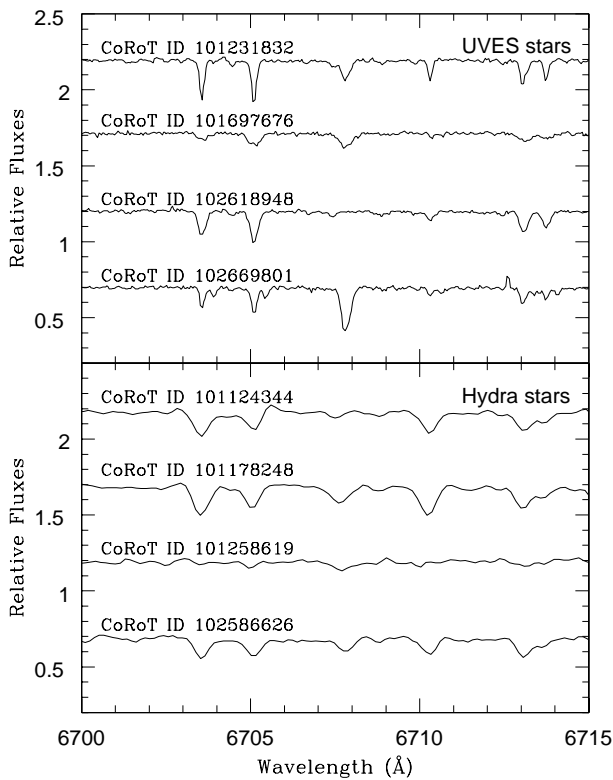}
   \caption{Some spectra of the stars contained in the sample. Spectra collected using the UVES spectrograph are 
   show in the \emph{upper panel}, whereas the \emph{lower panel} shows spectra collected using the Hydra spectrograph.}
              \label{fig:spectra}%
    \end{figure}

A complementary sample of $85$ stars was observed using the Hydra multifiber 
echelle spectrograph (henceforth \hydrastars), mounted on the Blanco 4m telescope at the Cerro Tololo 
Interamerican Observatory, located in Cerro Tololo, Chile. The filters $E5187$  ($509-525~nm$) and 
$E6757$  ($656-681$~nm) with a $200$~micron slit were used, allowing us to collect spectra with 
medium resolution ($R \approx 17,\!000$ and $15,\!000$, respectively) and signal-to-noise ratio $70<S/N<200$.
 The $E5187$ filter was chosen because it  covers a spectral region with five Fe II lines, whereas the 
$E6757$ was chosen because several Fe I lines  and a Li doublet (at $\approx 671 \, {\rm nm}$) are located in spectral 
window.  Also,  a few stars, with accurate previous measurements of \vsini and with FGK spectral types, were 
also observed with the same Hydra setups \citep{Melo_etal2001} to construct a calibration 
for the determination of \vsini for the Hydra stars in our sample.  Figure~\ref{fig:spectra} shows examples of 
these observations using both instruments.
 
 Table~\ref{TAB:OBSHYDRA} shows the setup used in Hydra observations. 
For the targets, we also compiled luminosity classes, $V$-band magnitudes, and color   indices from 
the CoRoT database\footnote{Available in \url{http://idoc-corot.ias.u-psud.fr/index.jsp}}, as well as 
$JHKs$ magnitudes from the Two-Micron All-Sky Survey (2MASS) database 
\citep{1996AJ....112.2168S}.  The main characteristics of the targets and the observation 
dates, and corresponding 
luminosity classes (from the CoRoT and 2MASS databases) are given in Tables~\ref{TAB:FHYDRA} and~\ref{TAB:OBSCALIB}.



%



The reduction of \uvesstars data was done using the standard UVES data reduction pipeline 
\citep{Ballester_etal2000}. 
Hydra data were reduced using the \emph{dohydra} task in IRAF.\footnote{IRAF is distributed by the National Optical 
Astronomy Observatories, which are operated by the Association of Universities for Research in Astronomy, Inc., under 
cooperative agreement with the National Science Foundation.}  
Both reductions follow the usual reduction steps  
(bias, flat-field, and background corrections, fiber order definition, wavelength calibration of the spectra
with Th-Ar lamp spectra, and extraction of the spectra). Then we use IRAF to normalize the spectra to a pseudocontinuum level and to 
bring the reduced spectra to the rest frame. Cosmic rays were extracted using the procedures described in
\citet{van_Dokkum2001}. 

\section{ Stellar properties}\label{CAP:DET}

\subsection{ Radial velocities \vrad} \label{CAP:RV}

\begin{figure}[!h]
   \centering
   \includegraphics[scale=0.8]{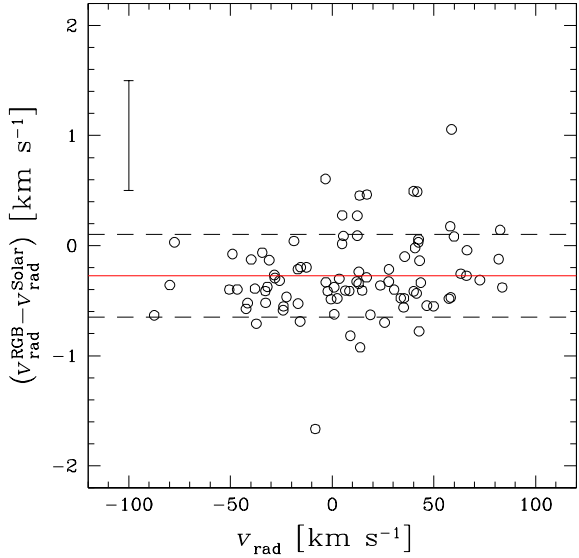}
   \caption{ Comparison between the \vrad  measurements for the Hydra sample obtained using different synthetic spectra. 
   The  $v_{\rm{rad}}^{RGB}-v_{\rm{rad}}^{Solar}$ represents the difference
   between the \vrad obtained when the Hydra spectra are  cross-correlated with synthetic spectra for a RGB star and the Sun, 
   respectively.  The red line represents
   the mean value of $v_{\rm{rad}}^{RGB}-v_{\rm{rad}}^{Solar}$ and its the standard deviation is presented using black dashed lines.  
   Also, the typical error in \vrad measurements for \hydrastars is shown using an error bar.}
              \label{fig:compRV}%
    \end{figure}

We obtained radial velocities \vrad  with the \emph{fxcor} task \citep{Tonry_Davis1979} in IRAF. 
Because the stars of the UVES sample present \teff close to the solar value, we cross-correlated the UVES spectra   
with a spectrum of the Sun \citep{2000vnia.book.....H}. We then converted the shifts 
into radial velocities of the stars, and we applied a  barycentric correction.   On the other hand, because the \hydrastars present a greater 
spread in temperatures and luminosity classes, the spectra were cross-correlated with  synthetic spectra of the Sun and an RGB star 
(\teff$=4000$~K, \logg$=1.0$~dex and \fe$=0.0$~dex) to compare differences in the determinations of radial velocities.
 We computed the spectra  with the Turbospectrum code \citep{Alvarez_Plez1998} and MARCS atmosphere models with solar 
abundances \citep{Gustafsson_etal2008}.  In Fig.~\ref{fig:compRV} small differences can be found between the  radial velocities  
derived using the synthetic spectra (averaging about $-0.27\pm0.37$~\kms).
 We opted to use the values found using the synthetic solar spectrum, and applied a  barycentric correction. The 
 typical errors in radial velocities for \hydrastars are lower than $0.5$~\kms.  On the other hand,  we also computed a synthetic solar 
 spectrum for the UVES spectral resolution, which we  used to obtain \vrad for the UVES sample, to 
 find systematic errors in the Hydra \vrad measured using synthetic spectra.
We found that our measurements of \vrad using synthetic spectra present a systematic difference of about $-0.75\pm1.66$~\kms with those
derived with observed solar spectrum, and is thus not significant.


%



\subsection{Rotation velocities \vsini}\label{CAP:VEL}

The \vsini measurements of our targets were computed using two procedures. For the case of 
\uvesstars, the \vsini  values were determined using the same procedure as in \citet{CantoMartins_etal2011}. Following these 
authors, the resulting spectra (taking into account the instrumental profile of UVES) are convolved with rotational profiles to
adjust the broadening observed in the iron lines  (profile fitting) located between $6700$ and $6715$~\AA. 

The \vsini values for the  \hydrastars were computed  (using the \emph{fxcor} task) with a cross-correlation function (CCF)  
especially calibrated for the Hydra  spectrograph. We then followed the same procedure decribed before for the .  
As is described in \citet{RecioBlanco_etal2002} and \citet{Lucatello_Graton2003}, the relation between \vsini 
and the  corrected width  $\sigma_{obs0}$ of the CCF is  

\begin{equation}\label{eq:CCF}
v \sin\,(i)\,= A \sqrt{(\sigma_{obs0})^2-(\sigma_0)^2} ~~\rm{km~s^{-1}},
\end{equation}

\noindent where $A$ and  $\sigma_0$ are the so-called coupling constant and the nonrotational contribution to the CCF width, respectively.
As mentioned in \citet{Melo_etal2001}, $\sigma_0$ depends on different broadening mechanisms (magnetic field, instrumental profile,
thermal broadening, etc.), which  is related to  object star and the template used,  but does not depend on rotation. 
 
Since the setup  $E5187$ presents the highest resolution in our Hydra observations, it was chosen to  determine the \vsini values. We used FGK 
stars with reliable \vsini determinations as templates and calibrators, which   were observed during our observing runs. The
\vsini values and photometry for these stars  are compiled in Table~\ref{TAB:OBSCALIB}.    

 The \emph{fxcor} task allowed us to obtain the uncorrected width of the CCF ($\sigma_{obs}$),  which has a contribution from the  
 template used ($\sigma_t$) in deriving the CCF. 
The $\sigma_t$ can be determined with an autocorrelation for each template, as is described in 
equation~4 of \citet{Lucatello_Graton2003}. For each star used as template we obtained  several spectra, which allowed us to avoid  the 
autocorrelation of the same  spectrum. The $\sigma_{obs}$ and  $\sigma_t$ are related with the corrected width  $\sigma_{obs0}$   through 
the following equation:

\begin{equation}
(\sigma_{obs0})^2=(\sigma_{obs})^2-(\sigma_t)^2.
\end{equation}



\begin{figure}[!h]
   \centering
   \includegraphics[scale=0.6]{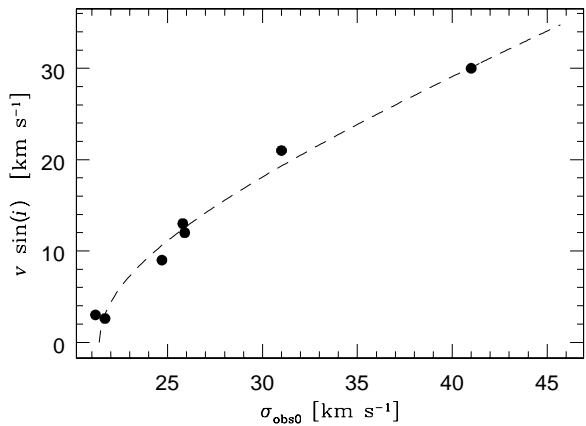}
   \caption{The CCF-\vsini calibration for the Hydra spectrograph. The dashed line represents the function found and black points
   represent the values for the calibrator stars, respectively.}
              \label{fig:CCF}%
    \end{figure}

The mean values of $\sigma_{obs}$ and  $\sigma_t$ for each template are listed in Table~\ref{tab:Template}, whereas the mean values of 
$\sigma_{obs}$ for each calibrator star  are listed in Table~\ref{tab:Calibrators}.  Finally, using a linear fit in the plane 
$[(\sigma_{obs0})^2,(v\sin(i))^2]$ 
the following relation between \vsini and $\sigma_{obs0}$ was obtained: $$v\sin (i)=(0.861\pm0.017)\sqrt{\sigma_{obs0}^2-(21.404\pm0.835)^2} ~~\rm{km~s^{-1}}.$$
 The errors in these coefficients are associated 
 with the errors in the slopes of the linear fit. 
 In Fig.~\ref{fig:CCF} we show the final calibration, which presents a good agreement with the
reference values of \vsini for calibrator stars.


%





These \vsini values were used  as reference to obtain new measurements of \vsini values using the setup $E6757$ and the 
same method described for \uvesstars. Note that  the setup $E6757$ contains the spectral region  between $6700$ and 
$6715$~\AA.  Small differences were found between \vsini values derived using both methods. Since we use
the profile fitting to derive lithium abundances  (as is described in section~\ref{CAP:FE}), the final values of \vsini 
for the \hydrastars are those derived from the profile fitting.





\subsection{Effective temperatures \teff, surface gravities \logg, iron abundance \fe, and microturbulence velocity \vmic}\label{method}

 Because our stellar sample is comprised of stars belonging to the field, it is important
 to have an estimation of their stellar parameters, which were used to avoid mistakes in the 
determination of the final parameters. In this sense, as a first step in the derivation of 
atmospheric parameters, we used the 2MASS $JHK$ near-infrared photometry and the CoRoT database 
mean values of $\langle V\rangle$ and $\langle (B-V)\rangle$ to  obtain a first  estimation of the  effective 
temperature for our sample. Specifically, we used the mean CoRoT color index $(B-V)$ and the 
calibrations of \citet{Flower1996} corrected by \citet{Torres2010} to calculate the photometric
effective temperature $T_{\rm{eff}}(B-V)$. In the same way, we calculated the photometric
effective temperature $T_{\rm{eff}}(J-K)$ using the 2MASS color index $(J-K)$, the CoRoT luminosity 
classes, and the calibrations of \citet{Alonso_etal1996,Alonso_etal1999}. 
We derived these temperatures  without reddening corrections 
(see section~\ref{CAP:RED} for detailed discussion). Also, we found
 errors at levels of $0.04$~mag  in the $(J-K)$ for the stars in our sample, which implies  
 errors at levels of $130$~K in  $T_{\rm{eff}}(J-K)$. No errors are informed in the CoRoT database for $(B-V)$. 
 
 We used the average between both photometric 
temperatures  as our initial estimation  for the  spectroscopic temperature. At the same time, we estimated the initial 
values of surface gravities \logg  using the CoRoT luminosity classes and 
interpolations in tables of infrared synthetic colors computed with ATLAS9 by R. Kurucz.\footnote{These 
tables are available on the Kurucz webpage \href{http://kurucz.harvard.edu/}{http://kurucz.harvard.edu/}.} 
  These initial estimates  can present important errors, produced by reddening and 
bad identification of CoRoT luminosity classes (see Fig.~8 in G10). In fact we found
differences of about $>500$~K between photometric and spectroscopic temperatures, which implies
high rates of extinction in the CoRoT fields (see Section~\ref{CAP:RED}). For this reason 
we stress that these initial values were used only as a starting point to obtain the final parameters and,
in any case, they constrained the searching of spectroscopic temperatures.

\begin{figure}[!h]
   \centering
   \includegraphics[scale=0.8]{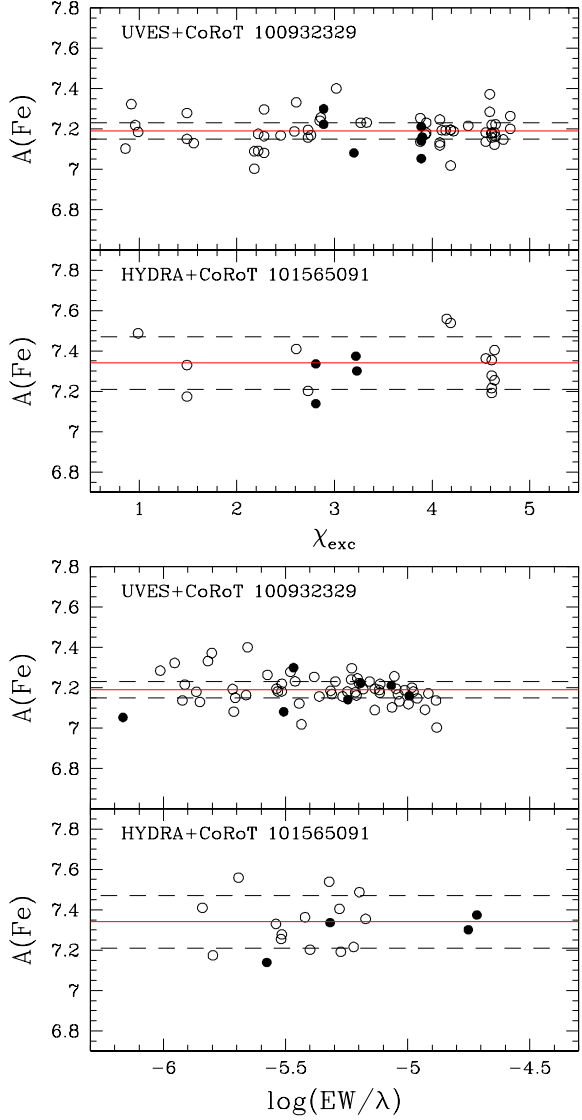}
   \caption{\emph{Upper panels}: ionization equilibrium for the targets 
   $100932329$ (UVES star) and  $101565091$ (Hydra star).
   \emph{Lower panels}: equilibrium of the $A(\rm{Fe})$s and their $EW$s. Open and filled circles represent the abundances of FeI ($A(\rm{FeI})$)
   and FeII ($A(\rm{FeII})$), respectively. Red and black dashed lines represent the 
   mean value of  $A(\rm{Fe})$ and standard deviation, respectively.}
              \label{fig:equilibrium}%
    \end{figure}

 We determined final values of the atmospheric parameters and their respective errors  using the Turbospectrum code 
 and MARCS atmosphere models with solar abundances. Solar abundances were taken from 
\citet{Asplund_etal2005}, and the collisional damping treatment was performed based on the work of Barklem and 
co-workers \citep{Barklem_etal2000a,Barklem_etal2000b,Barklem_Piskunov2003,Barklem_AspelundJ2005}. To compute synthetic spectra 
with the Turbospectrum code, we took  atomic (see below) and molecular line lists into account, including TiO 
\citep{Plez1998}, VO \citep{Alvarez_Plez1998}, and CN and CH \citep{Hill_etal2002}. The Turbospectrum code uses equivalent 
widths ($EW$)  to compute abundances $A(\rm{Fe})$ corresponding to the different Fe lines\footnote{We use the FeI 
and FeII abundances defined as $A(\rm{FeI})=\log_{10} \left(\frac{N_{\rm{FeI}}}{N_{\rm{H}}}\right)$ and $A(\rm{FeII})=\log_{10} \left(\frac{N_{\rm{FeII}}}{N_{\rm{H}}}\right)$, 
respectively.}. 
 The list of Fe lines used  was compiled and corrected by 
\citet{CantoMartins_etal2011}  (see their Table~6).  This list is composed of  $91$   and $14$ lines of Fe I and II, respectively. 
There are differences in the number of iron 
lines used to characterize the stars because our sample was observed using two instruments and different setups. In particular, for UVES
stars, we used all lines compiled by \citet{CantoMartins_etal2011}, whereas for the  Hydra stars $33$ of the lines ($28$ FeI lines and $5$ FeII lines) 
in the intervals $5090-5250$~\AA~ and $6560-6810$~\AA~ were
used. We measured $EW$ values  with the DAOSPEC code \citep{Stetson_Pancino2008}.
Using excitation equilibrium for the FeI abundances, FeI/FeII ionization 
equilibrium, and the  equilibrium of the $A(\rm{Fe})$ values and their respective $EW$ values,  we can derive effective 
temperatures \teff, surface gravities \logg, and microturbulence velocities \vmic, respectively.  Starting from the 
photometric parameters, we ran the Turbospectrum code iteratively using MARCS atmosphere models with different parameters 
to find the three equilibria, thus defining  the final parameters. 
Figure~\ref{fig:equilibrium} presents an example of physical and chemical parameter  determinations using the 
equilibria (slopes equal to zero in the planes shown in this figure) for both a UVES star and a Hydra star.


%




 The errors in \teff and \vmic  are obtained from the errors in the slopes that define the equilibria described above.  We change one of these 
parameters, keeping the others fixed, and compute a new atmospheric model. We found the error when the slope of the new 
fit became equal to its respective slope error. The error in \fe  is equal to slope error in excitation equilibrium for the 
FeI abundance. Finally, the error in \logg is found when the difference between $A(\rm{FeI})$  and $A(\rm{FeII})$ is equal to the square root 
of the  sum of the squares of the errors in $A(\rm{FeI})$  and  $A(\rm{FeII})$.

\subsection{The Li abundance determinations}
\begin{figure}[!h]
   \centering
   \includegraphics[scale=0.6]{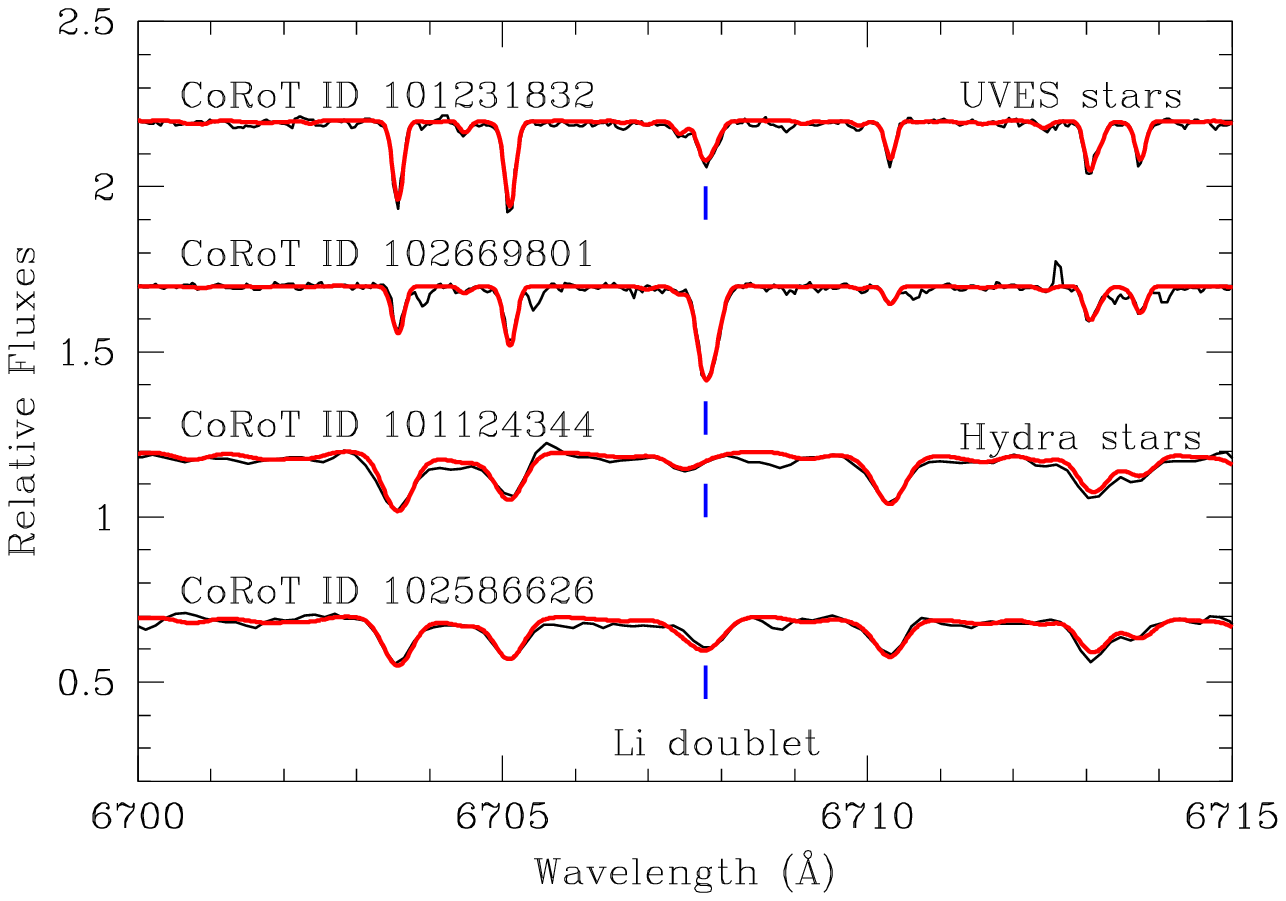}
   \caption{ The profile fitting of the Li doublet at $\sim6708$~\AA~ is shown for two UVES stars (CoRoT ID $101231832$ and $102669801$) and two
   Hydra stars (CoRoT ID $101124344$ and $102586626$). The observed spectra are presented using black lines, whereas the synthetic spectra  are 
   shown in red. The position of the  Li doublet    is indicated with a blue vertical line.}
              \label{fig:li}%
    \end{figure}

The $\rm{Li}$ abundances \li\footnote{Here the $\rm{Li}$ abundance is defined as $A(\rm{Li})=\log_{10}\left(\frac{N_{\rm{Li}}}{N_{\rm{H}}}\right)-12.00$} for 
\emph{UVES} and \hydrastars were calculated by fitting  the observed profile with a synthetic profile  of the $\rm{Li}$  doublet located at 
$\sim6708$~\AA.  The synthetic spectrum was computed using the Turbospectrum code. Figure~\ref{fig:li} shows four examples showing the method used
to determine the  $A(\rm{Li})$.


%




   \begin{figure*}[!h]
   \centering
   \includegraphics[scale=0.9]{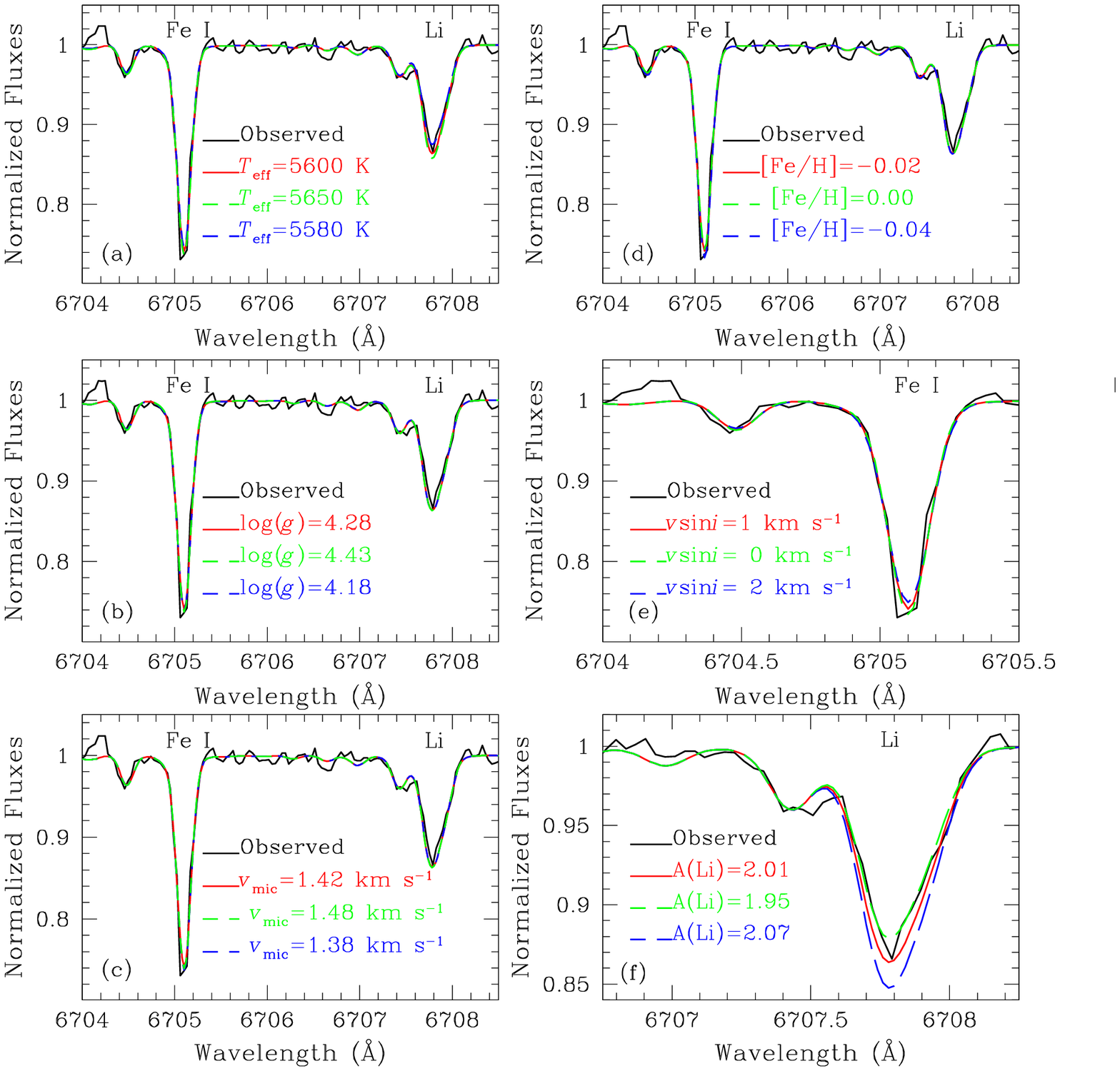}
   \caption{   As in Fig.~\ref{fig:li}, for the CoRoT ID $101231832$ the profile fitting of an Fe I line  and the  Li doublet 
   are presented. The panels (a), (b), (c), and (d) show  how the errors in each stellar parameter impact  the computed synthetic spectra, which were used to measure \vsini (panel e) and $A(\rm{Li})$ (panel f), and their respetive errors. For all panels, the observed spectrum is presented using a black line, whereas the synthetic spectrum, computed with the derived stellar parameters, is presented using a red line. The dashed green and blue lines represent the computed synthetic spectra with errors for a given stellar parameter.   }
              \label{fig:lizoom}%
    \end{figure*}
%



 The error $\sigma_{A({\rm Li})} $ in this abundance is related to the errors in the physical and chemical parameters; more specifically, the magnitude of 
 this error is directly related to  the error in \teff, and the errors in the other stellar parameters, such as \logg, \vmic, and \vsini,  produce 
 minor effects in the measurements of $A({\rm Li})$.   We determined new lithium abundances using synthetic spectra reflecting 
 the errors in the four stellar parameters described above, which we called $A(\rm{Li})^{j}_{ER}$. Then, the final error $\sigma_{A({\rm Li})}$ is equal to the square 
 root of the  sum of the squares of the difference between \li and $A(\rm{Li})^{j}_{ER}$, as is described in the following equation:
 
\begin{equation}
        \sigma_{A({\rm Li})}=\sqrt{\sum_{\rm{j=1}}^{\rm{j=4}}\left(A({\rm Li})-A(\rm{Li})^{j}_{ER}\right)^2}
.\end{equation}
 
  The Fig.~\ref{fig:lizoom} shows how the errors in the stellar parameters impact in the $A({\rm Li})$. In this context, the  measurements  of 
 $A({\rm Li})$  for \hydrastars present errors higher than those found for the . Nevertheless, we should be cautious with 
 the Hydra data due, in particular, to the  spectral resolution ($R\sim15,\!000$) associated with the observations.

\section{Results}\label{CAP:RESULTS}

\begin{figure}[!h]
   \centering
   \includegraphics[scale=0.8]{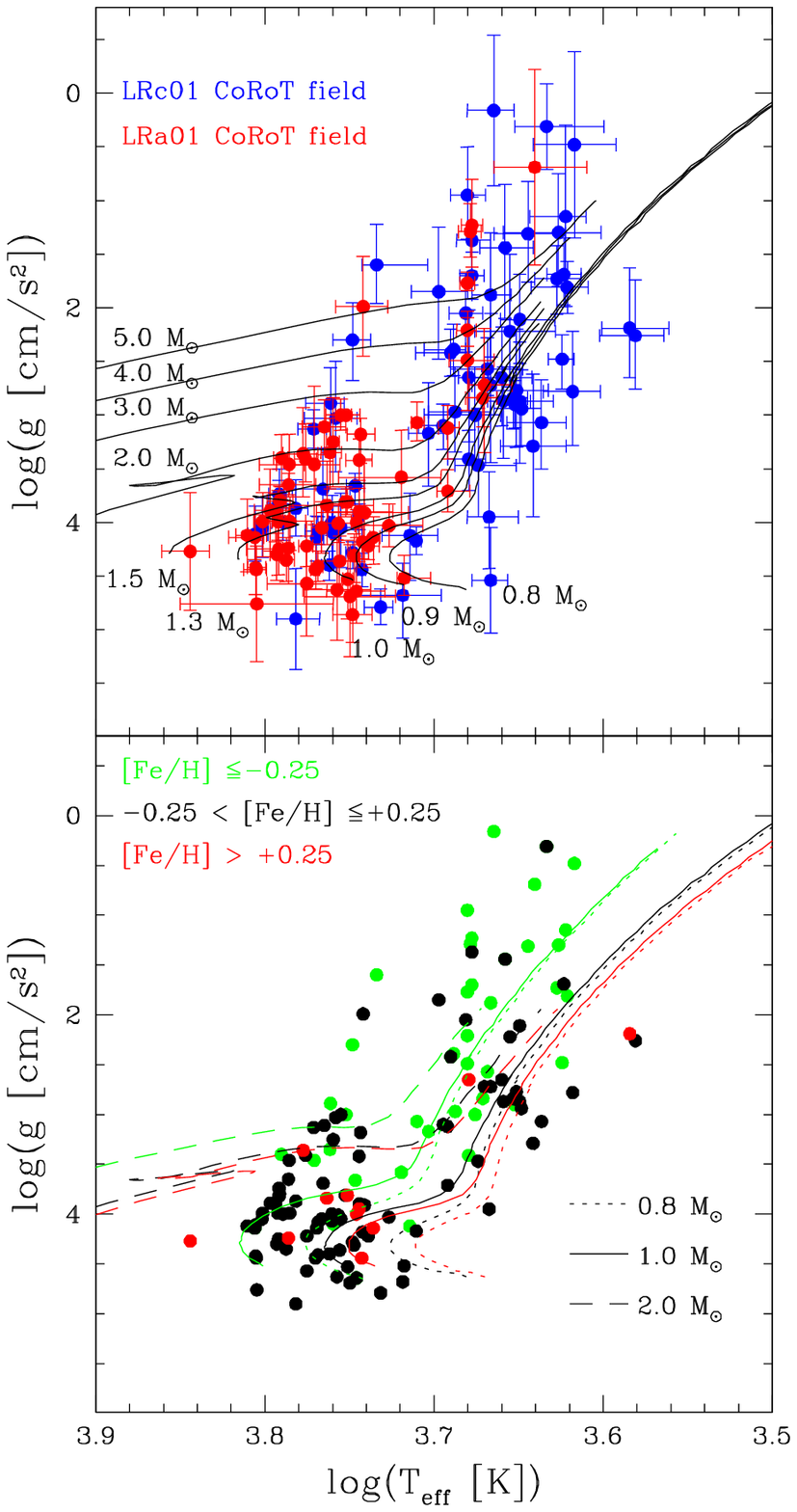}
   \caption{\emph{Upper panel}: HR diagram of the stellar 
   sample segregated by Galactic direction (center/anticenter). Blue and red dots represent  stars in the \cen ~and \anti ~fields, respectively.
    Evolutionary tracks for $Z = 0.019$, from Girardi et al. (2000), are shown for stellar masses between 0.8 and 5~M$_\odot$.
  \emph{Lower panel}: HR diagram of the stellar sample segregated by the 
   iron abundance $\rm{[Fe/H]}$. Green, black, and red points represent stars with  \fe$\leq -0.25$, $-0.25<$\fe$\leq +0.25$ and \fe$\geq +0.25$, 
respectively.  Evolutionary tracks for $Z = 0.004$, $Z = 0.019$, and $Z = 0.030$, from Girardi et al. (2000), are  presented with 
lines in green, black, and red, respectively. In this panel, only the following stellar mass values are represented: $0.8~M_\odot$ 
(dot lines), $1.0~M_\odot$ (solid lines) and $2.0~M_\odot$ (dash lines).  For clarity, in both panels we plot only the evolutionary tracks until to the RGB stage.
 }
              \label{fig:HR}%
    \end{figure}

The computed stellar parameters, including rotational velocities \vsini and lithium abundances, for the present stellar sample 
are listed in Table~\ref{TAB:ALLPARAMETERS}. Figure~\ref{fig:HR} presents the corresponding Hertzsprung-Russell (HR) diagram, 
with the stars segregated by their Galactic locations (CoRoT center/anticenter) and iron abundances. In the bottom panel, stars 
are divided in three different groups using their \fe values.\footnote{\fe is 
calculated as in \citet{CantoMartins_etal2011} using a solar iron abundance $A(\rm{Fe})_{\odot}=7.\!49$~dex}
Errors in the parameters are also included in these panels. The magnitude of these errors is 
linked to the quality of the spectra, including spectral resolution $R$ and $S/N$, and intrinsic effects 
of the stellar surfaces (i.e., high \vsini, molecular bands in cool stars, etc.). Figure~\ref{fig:HR} shows that the 
present sample is comprised of stars in different evolutionary stages, ranging from the main sequence (MS) to the red giant branch (RGB). 


%


We used the evolutionary tracks of \citet{Girardi_etal2000} for different 
masses and metallicities\footnote{For the different \fe groups, we used  evolutionary 
tracks with a representative $Z$. Specifically, we used metallic abundances $Z=0.004$, $0.019$ and $0.030$ 
for the group with \fe$\leq -0.25$, $-0.25<$\fe$\leq +0.25$ and \fe$\geq +0.25$, 
respectively.} to identify the evolutionary stages of the stars in our sample. For this purpose, we identified
the turn-off and the base of the RGB from the evolutionary tracks for each $Z$ to define 
the MS, subgiant branch (SGB), and RGB regions. The results of this classification are listed in Table~\ref{TAB:ALLPARAMETERS}. 
  Mean values for the stellar  parameters and their respectives standard deviations corresponding to different CoRoT fields 
 and evolutionary stages are given in  Table~\ref{tab:meanparam}. 


%
   \begin{figure}[!h]
   \centering
   \includegraphics[scale=0.8]{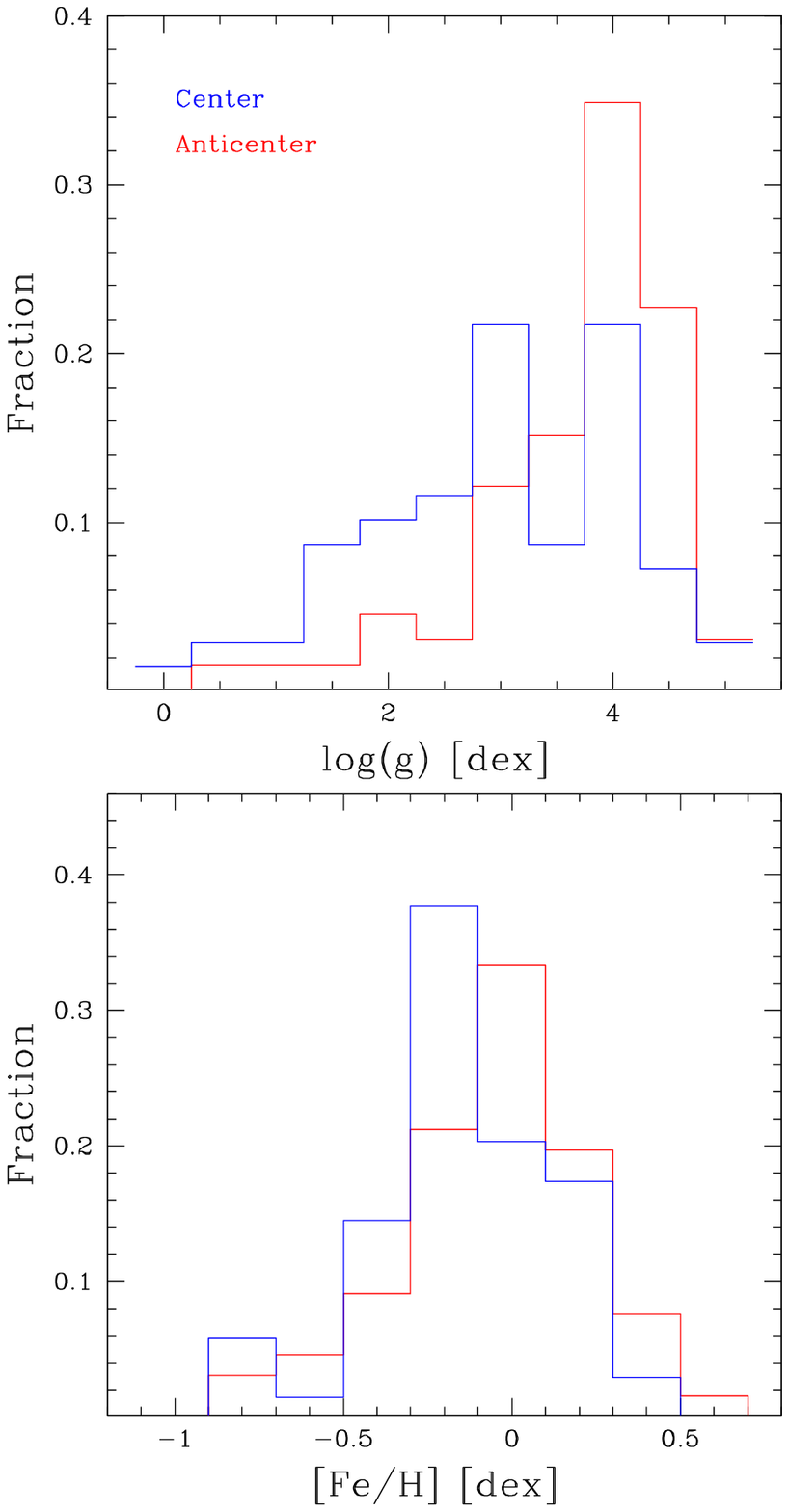}
   \caption{\emph{Upper panel}: histogram of the surface gravities \logg in the CoRoT fields. \emph{Lower panel}: 
   histograms of the iron abundances \fe    for the different evolutionary stages. A relation between \fe and 
   evolutionary status is clearly present for the stars in our sample.}
              \label{fig:histologgFe}%
    \end{figure}
%


 Figure~\ref{fig:HR} also shows that the stellar evolutionary distribution for LRc01 and LRa01 CoRoT fields agrees 
 with \citet{Deleuil_etal2009}.  Furthermore, the relation between the color and evolutionary status  is linked to the  stellar metallicity.  In fact, as we can see in Fig.~\ref{fig:histologgFe} and  
Table~\ref{tab:meanparam}, the
distributions of \logg for both fields are different and these differences correlate with the \fe distributions. Most \anti stars have
\logg$\geq3.75$~dex ($\sim 64\%$), whereas the \cen stars are spread along a large interval of \logg and  most of 
them have \logg$\leq3.75$~dex. Combining this with the fact that the peaks of the \fe distributions  and  their mean values 
(Table~~\ref{tab:meanparam}) are different  {\bf from} one another, we can see
that there is a relation between temperature, surface gravity, and evolutionary stage in the CoRoT fields considered here.
  While one might be tempted to  
associate these differences  to selection effects, we note that  these results agree with 
the recent spectroscopic survey of the CoRoT fields presented by G10. This point
is discussed further in Sec.~\ref{CAP:FE}.

\subsection{Comparison with previous results}

\begin{figure*}[!h]
   \centering
   \includegraphics[scale=0.8]{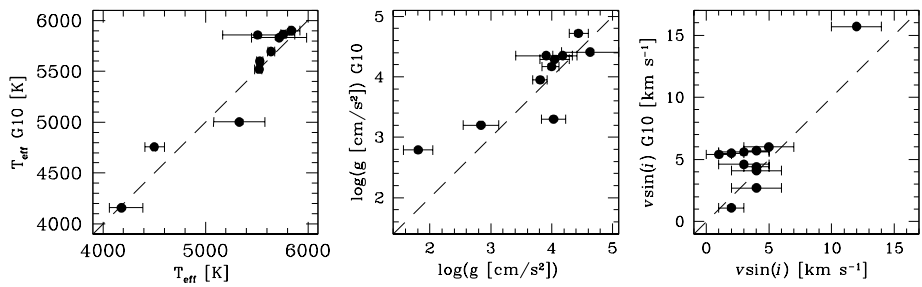}
   \caption{Comparison between our results and the results of G10 for  stars in common.  Some differences can be noted, however, overall
our result agree with G10.
 }
              \label{fig:common}%
    \end{figure*}

To verify our results, we compared them with the results presented in G10. Only 11 stars of our sample
were also analyzed by G10. These stars are listed in Table~\ref{TAB:common}. In Fig.~\ref{fig:common}
we plot the comparison between our results and the  G10 findings.  Our results agree with 
the survey of G10, with \teff, \logg and \vsini  presenting only small differences for most of the stars. However, star $101538522$ 
presents a large
dispersion in the \logg values, which may be explained because of the quality of the data of G10 for a RGB star ($S/N\sim23$).  
We cannot  directly compare their derived abundances with ours since G10 report only the global metallicity ($\rm{[M/H]}$), whereas here we present
the iron abundance (\fe).


%


\subsection{Radial and rotational velocities}\label{CAP:RADIAL_VSINI}

\begin{figure}[!h]
   \centering
   \includegraphics[scale=0.8]{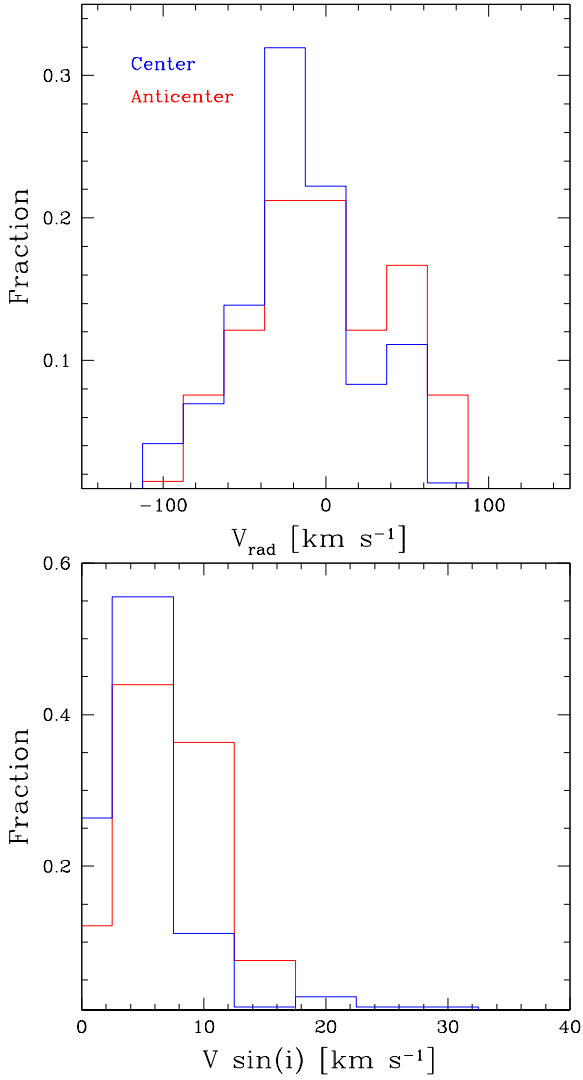}
   \caption{\emph{Upper panel}: histograms of the radial velocity \vrad for the \cen ~(blue line) and \anti ~(red line) fields.
\emph{Lower panel}: histograms of the rotational velocity \vsini for the \cen ~(blue line) and \anti ~(red line) fields.}
              \label{fig:RV}%
    \end{figure}

Figure~\ref{fig:RV} shows the distribution of the measurements of barycentric radial velocity \vrad  and rotational velocity 
\vsini listed in  Tables~\ref{TAB:ALLPARAMETERS} and \ref{tab:meanparam}.  Small, but significant, differences are 
observed in the  \vrad distribution for stars located in the Galactic center and anticenter directions. The percentages 
of stars with \vrad$\leq0$~\kms associated with the Galactic center and anticenter directions are $64~\%$ and $50~\%$, respectively.  
In spite of the incompleteness of the sample, this behavior agrees with the \vrad distribution determined by G10 (see their
Fig.~2).

\begin{figure}[!h]
   \centering
   \includegraphics[scale=0.8]{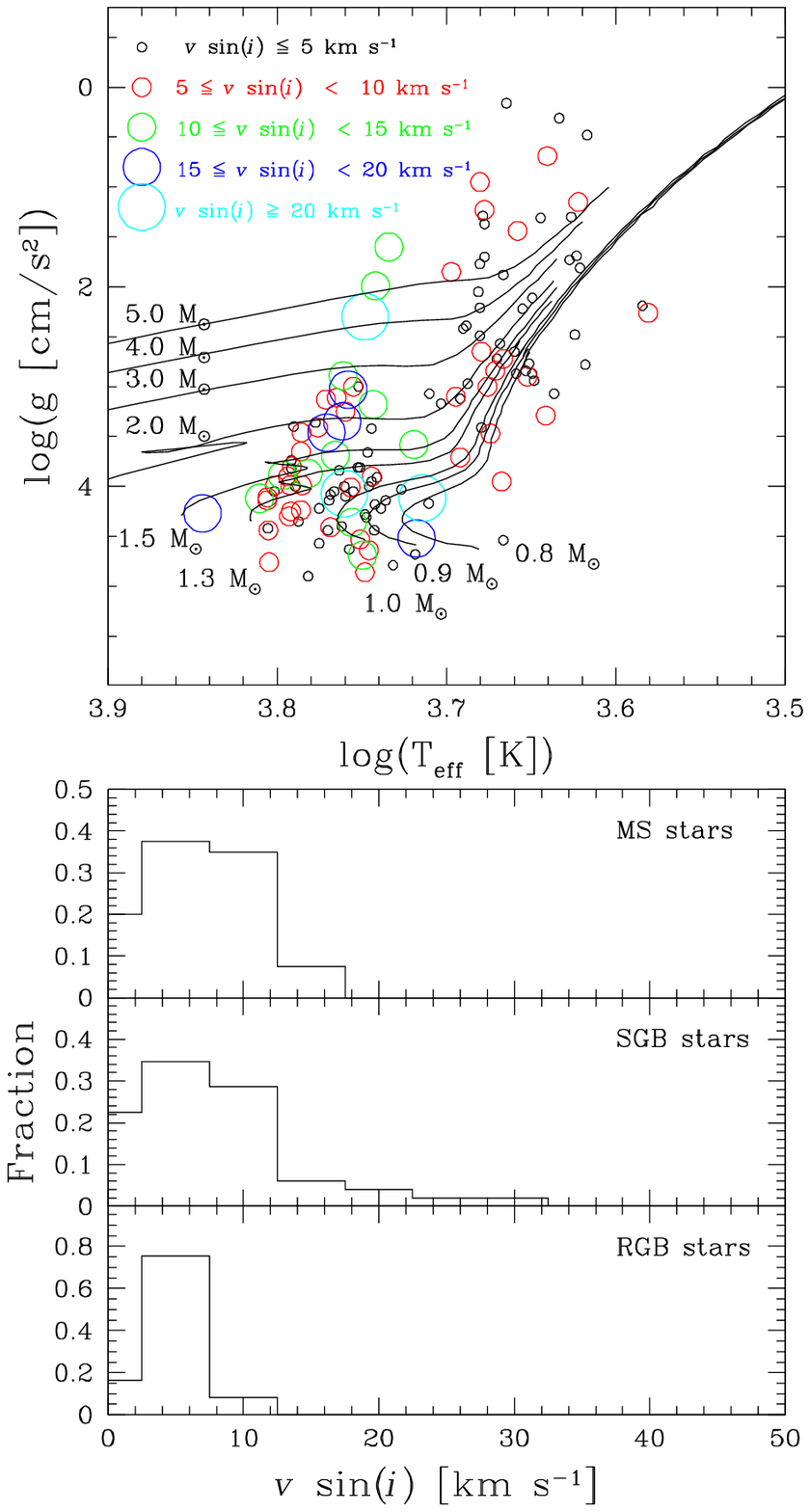}
   \caption{\emph{Upper panel}: distribution of \vsini for the stars in our sample along the HR diagram. Symbol size is proportional to the 
   \vsini value. Evolutionary tracks are defined as in Fig.~\ref{fig:HR}.  \emph{Lower panels}:  histograms of the rotational velocity 
   \vsini for the different evolutionary stages.}
              \label{fig:HRVSINI}%
    \end{figure}

In Fig.~\ref{fig:RV},  small differences are also observed in the  distribution of \vsini values. In fact, 
 $68~\%$  and $42~\%$ of stars in the \cen and \anti fields present \vsini$\leq5$~\kms, respectively.  The difference 
 disapears quickly for stars  with  \vsini$\leq10$~\kms, where the percentage for both fields are similar 
 ($89~\%$ and $83~\%$ for \cen and \anti fields, respectively). A  few high rotation values can be noticed among stars 
 in the Galactic center region While we caution that these distributions are affected to some degree by incompleteness, 
 the observed behavior of the \vsini distribution in particular does follow the behavior expected for FGK stars 
 \citep{Soderblom_etal1983,deMedeiros_etal1996,Nordstrom_etal2004}.  In fact, as we can see from Figure~\ref{fig:HRVSINI}, 
 which displays the individual \vsini~ values versus \teff, the rotational behavior for  stars of the present stellar 
 sample is rather well in agreement with the well-established behavior of rotation for stars evolving from the MS to red 
 giant stages.  Essentially, stars in the MS exhibit a wide range of rotational velocity values, which are related 
 with the stellar masses and \teff. For these stars, the measured values of  \vsini ranges from a few km/s to about 100 
 times the solar rotation rate, whereas the stars along the RGB are typically slow rotators, except for a few 
 unusual cases presenting moderate to rapid rotation \citep{Cortes_etal2009,Carney_etal2008,Carney_etal2003,deMedeiros_etal1996,demedeiros1990}.



%



%

\subsection{Photometric temperatures and reddening}\label{CAP:RED}

It is possible 
to evaluate the reddening effects along the two different Galactic directions by
comparing the initial photometric temperatures, derived without taking reddening 
into account, and the final, spectroscopically-derived, and presumably 
reddening-insensitive temperatures.
This analysis allows one to 
establish how the determination of physical parameters from photometric data  
is affected by neglecting reddening effects, to evaluate the 
error budget brought about by reddening, and also to check the presence of possible 
reddening gradients in the CoRoT  windows.
In this sense, the present $E(B-V)$ estimates for individual stars may assist 
follow-up programs of specific groups of stars, including for instance 
solar analogs and  solar twins.


   \begin{figure}[!h]
   \centering
   \includegraphics[scale=0.8]{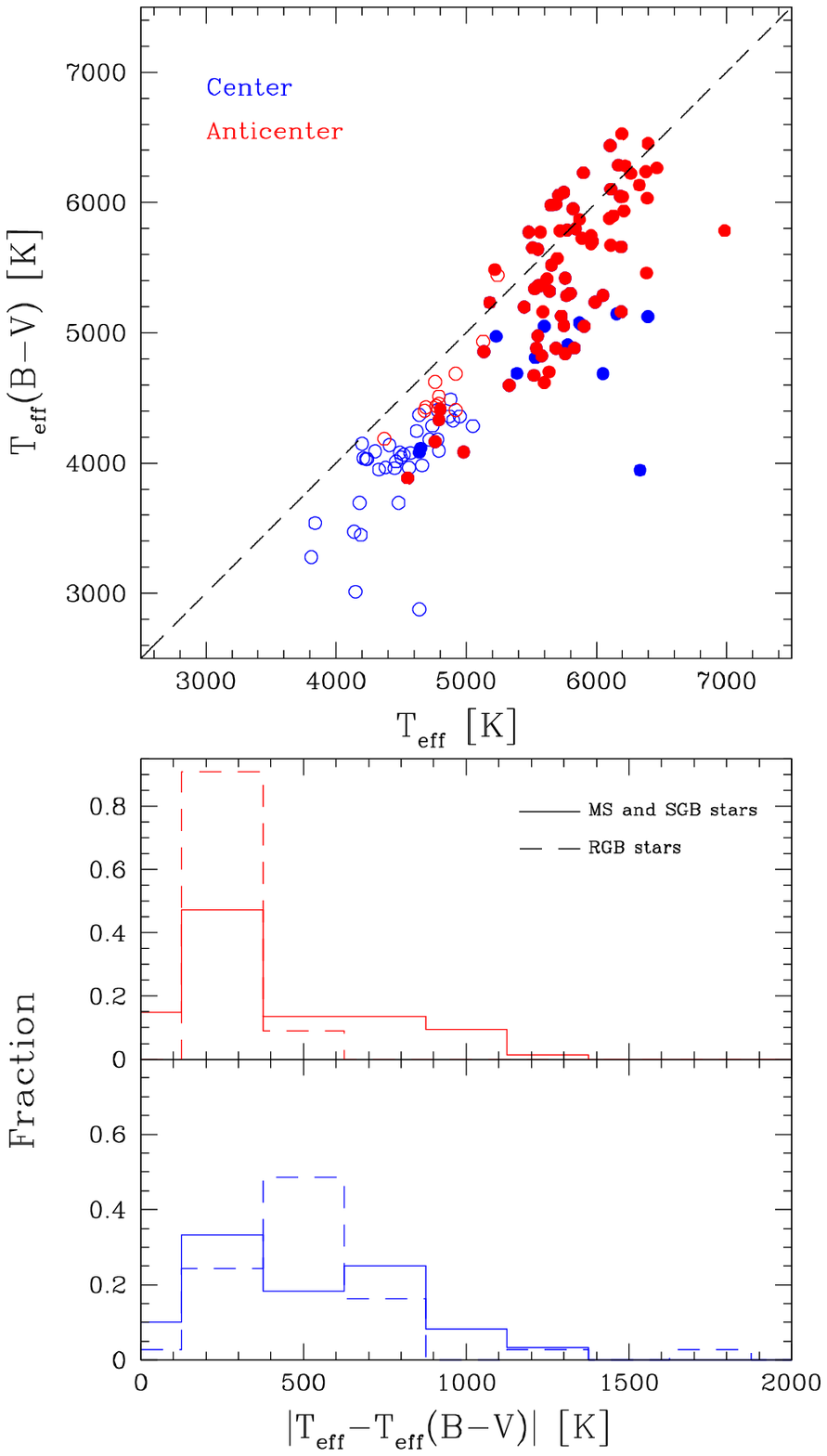}
   \caption{\emph{Upper panel}: differences between photometric $T_{\rm{eff}}(B-V)$ and spectroscopic
   \teff for  stars in the \cen ~(open and filled blue dots) and \anti ~(filled and open red dots)  CoRoT fields.
   Filled circles represent MS and SGB stars, whereas open circles represent RGB stars.
   \emph{Lower panels}: histograms of the difference between photometric $T_{\rm{eff}}(B-V)$ and spectroscopic
   \teff for the stars in the \cen ~(blue line) and \anti ~(red line)  CoRoT fields.
   }
              \label{fig:TBmV}%
    \end{figure}
%



   \begin{figure}[!h]
   \centering
   \includegraphics[scale=0.8]{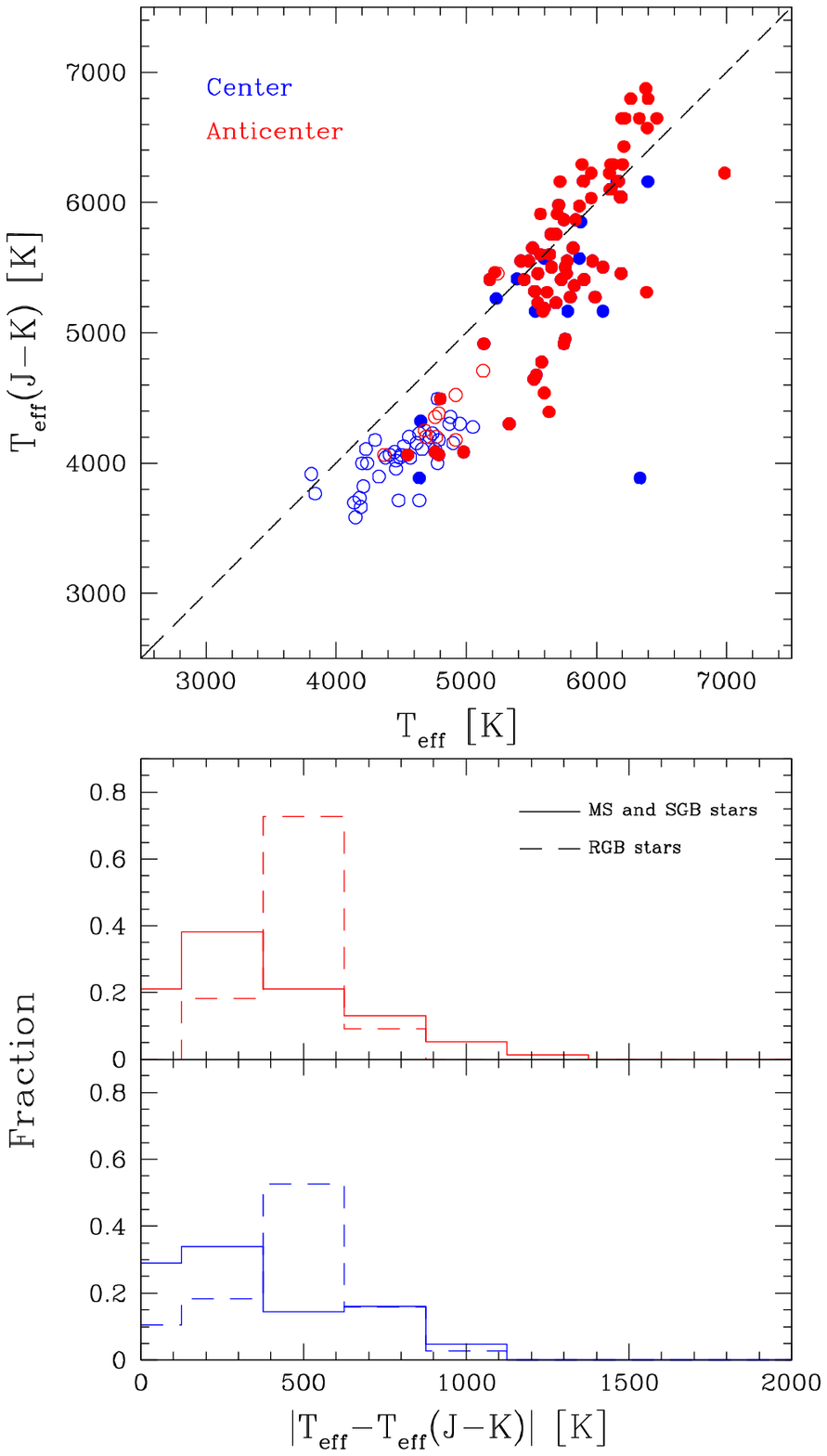}
  \caption{\emph{Upper panel}: differences between photometric $T_{\rm{eff}}(J-K)$ and spectroscopic
   \teff for the stars in the \cen ~(open and filled blue dots) and \anti ~(filled and open red dots)  CoRoT fields.
   Filled circles represent MS and SGB stars, whereas open circles represent RGB stars.
    \emph{Lower panels}: histograms of the difference between photometric $T_{\rm{eff}}(J-K)$ and spectroscopic
   \teff for the stars in the \cen ~(blue line) and \anti ~(red line)  CoRoT fields.
   }
              \label{fig:TJmK}%
    \end{figure}
%



To accomplish this goal, we compare the photometric temperatures $T_{\rm{eff}}(B-V)$  
and $T_{\rm{eff}}(J-K)$ and those derived from our spectroscopic analysis, $T_{\rm{eff}}$. 
 We determined the photometric temperatures, at the beginning, using the luminosity class 
from the CoRoTSky database. For some stars, however, the physical parameters provide a new classification 
of luminosity class for which new photometric temperatures were  computed using this information. The
photometric temperatures are listed in Table~\ref{TAB:ALLPARAMETERS}, including  those derived using a 
new luminosity class.  Finally, in Figs.~\ref{fig:TBmV} and \ref{fig:TJmK}, we present the comparisons 
between our spectroscopic $T_{\rm{eff}}$ values and the photometric estimations $T_{\rm{eff}}(B-V)$ and $T_{\rm{eff}}(J-K)$.

Figure~\ref{fig:TBmV} shows that the stars in CoRoT run \cen ~present larger differences  between $T_{\rm{eff}}(B-V)$ and
$T_{\rm{eff}}$ than those in run \anti, which can be expected because of the higher extinction levels in the Galactic center
direction. In fact, for this color index, (B--V), the percentages of CoRoT stars presenting differences up to $200$, $500$, and $800$~K
in the Galactic center direction are $6~\%$, $42~\%$, and $81~\%$, respectively. The percentages of CoRoT stars presenting the 
same temperature differences in the Galactic anticenter direction are $38~\%$, $86~\%$, and $92~\%$, respectively. To obtain a 
reddening estimation for the LRc01 and LRa01 CoRoT fields, we used the calibration of \citet{Flower1996}(corrected by
Torres 2010), which give us  $T_{\rm{eff}}(B-V)$ and \teff, assuming the latter (spectroscopically derived) as being the actual value. This
assumption is valid because \teff is not affected by reddening. As such, for a star  with a solar value
$(B-V)_{\odot}$,  we estimated reddening  levels ($E(B-V)$)  of about  $0.06$, $0.17$, and $0.30$, for the differences of
$200$, $500$, and $800$~K, respectively. Similarly, for an RGB star with a $(B-V)_0\sim1.6$, these differences in 
temperature represent reddening levels of about $E(B-V)\sim$ $0.07$, $0.16$, and $0.20$, respectively. 

 We also derived $E(B-V)$ values for  individual targets using their values of  $T_{\rm{eff}}(B-V)$ and \teff of 
Table~\ref{TAB:ALLPARAMETERS}. These values are given in Table~\ref{TAB:ALLPARAMETERS}. Table~\ref{tab:meanparam} shows the mean values $<E(B-V)>$. Then, for all evolutionary 
stages, $<E(B-V)>~\sim0.27\pm0.14$ and $0.12\pm0.11$ for the \cen and \anti fields, respectively. These mean values are 
somewhat influenced by the  distributions of temperature and   evolutionary stage in each field. Then we computed mean 
values $<E(B-V)>$ for each evolutionary stage for both fields. For the \cen field, $E<(B-V)>$ levels are of 
$0.36\pm0.22$, $0.25\pm0.11$, and $0.25\pm0.12$ for MS, SGB and RGB stars, respectively, whereas for the 
\anti field, $<E(B-V)>$ levels are of $0.11\pm0.12$, $0.12\pm0.13$, and $0.13\pm0.06$ for MS, SGB, and RGB stars, respectively.

 
On the other hand, there are also differences when the $T_{\rm{eff}}(J-K)$ values are
compared with $T_{\rm{eff}}$ in  both CoRoT fields.  The percentages of CoRoT stars presenting differences up to $200$, $500$, and $800$~K
in the Galactic center direction are $15~\%$, $61~\%$, and $90~\%$, respectively. The percentages of CoRoT stars presenting 
the same temperature differences in the Galactic anticenter direction are
$38~\%$, $80~\%$, and $91~\%$, respectively. 

The reddening levels for both CoRoT fields were determined using the relations of \citet{Alonso_etal1996,Alonso_etal1999}, which provide $T_{\rm{eff}}(J-K)$ and  \teff, again assuming that the latter provides the correct value. As such, for a star with a solar value
$(J-K)_{\odot}$, reddening levels are of about $E(J-K)\sim$ $0.04$, $0.09$, and $0.16$ for differences up to $200$, $500$, and $800~K$, respectively.
Similarly, for a RGB star with a $(J-K)\sim1.0$ these 
differences in temperature imply a reddening of about $E(J-K)\sim$ $0.13$, $0.37$, and $0.71$.

 Similar to $E(B-V)$, we also derived $E(J-K)$ values for  individual targets using their values of  $T_{\rm{eff}}(B-V)$ and \teff of 
Table~\ref{TAB:ALLPARAMETERS}. These values are given in Table~\ref{TAB:ALLPARAMETERS}.  Table~\ref{tab:meanparam} shows mean values $<E(J-K)>$. Then, for all 
evolutionary stages, $<E(J-K)>~\sim0.14\pm0.10$ and $0.10\pm0.09$ for the \cen and \anti, respectively. For 
the \cen field, $<E(J-K)>$ levels are of $0.10\pm0.15$, $0.11\pm0.07$, and $0.17\pm0.08$ for MS, SGB, and RGB 
stars, respectively, whereas for the \anti field $<E(J-K)>$ levels are of $0.13\pm0.10$, $0.07\pm0.09$, and 
$0.14\pm0.04$ for MS, SGB, and RGB stars, respectively.

When we compare $<E(B-V)>$ with $<E(J-K)>$ for each field or the evolutionary stages in each field, typically 
they do not agree with one another. However,
the dispersion in these mean values is very high, which could explain this discrepancy. 

In contrast, the mean reddening values show that the \cen field is more affected by reddening than the \anti field. 
In fact, we used the reddening maps of \citet{Schlafly_etal2011} and \citet{Schlegel_etal1998}, and the Galactic 
Extinction Calculator\footnote{Available in the NED web page \url{http://ned.ipac.caltech.edu/forms/calculator.html}} 
of the NASA/IPAC  Extragalactic Database (NED) to obtain mean values of reddening of the \cen and \anti fields. However, 
NED only provides reliable reddening values for the \cen field. The reddening values of \citet{Schlafly_etal2011} and 
\citet{Schlegel_etal1998} agree with the present work. Specifically, for the \cen  field \citet{Schlafly_etal2011} 
give reddening values $E(B-V)\sim~0.28$ and $E(J-K)\sim~0.13$, whereas \citet{Schlegel_etal1998} give  $E(B-V)\sim~0.30$ and $E(J-K)\sim~0.16$.

\begin{figure}[!h]
   \centering
   \includegraphics[scale=0.8]{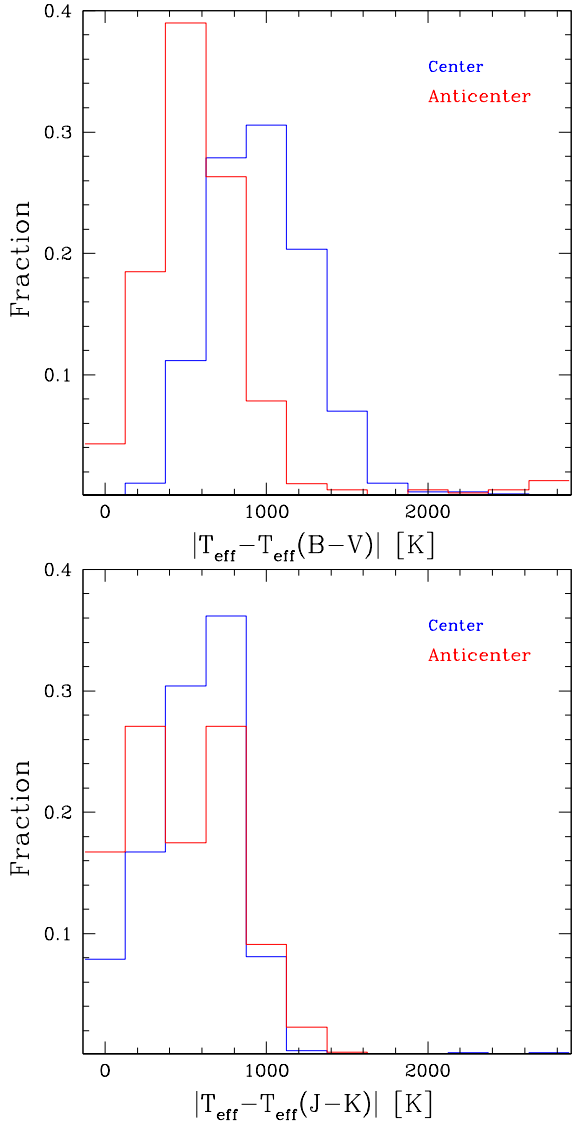}
  \caption{\emph{Upper panel}: histograms of the difference between photometric $T_{\rm{eff}}(B-V)$ and spectroscopic
   \teff for the sample of \citet{Gazzano_etal2010}. 
   \emph{Lower panel}: histograms of the difference between photometric $T_{\rm{eff}}(J-K)$ and spectroscopic
   \teff for the sample of \citet{Gazzano_etal2010}.}
      
          \label{FIG:GTJmK}%
    \end{figure}

To make a comparison, we also obtained the color indices $(B-V)$ and $(J-K)$ for the stellar sample of
G10. Again, we computed the $T_{\rm{eff}}(B-V)$ and $T_{\rm{eff}}(J-K)$ values for the stars for 
which those authors spectroscopically derived atmospheric parameters, and we obtained the differences between these temperatures and the
spectroscopic temperatures. Histograms showing these differences for the stars in the \cen ~and \anti ~fields are presented
in Fig.~\ref{FIG:GTJmK}. There are some differences between these distribution and the corresponding 
histograms derived in Figs.~\ref{fig:TBmV} and \ref{fig:TJmK}.  Compared to our sample, the G10 sample presents a greater 
proportion of highly-reddened stars in the Galactic center direction. 
 In fact, for the $(B-V)$ color, the percentages of CoRoT stars presenting differences up to $200$, $500$, and $800$~K
in the Galactic center direction are $0~\%$, $3~\%$, and $30~\%$, respectively. The percentages of CoRoT stars presenting 
the same temperature differences in the Galactic anticenter direction are $3~\%$, $41~\%$, and $80~\%$, respectively.


This difference can probably be explained by the fact that the relative number of stars in the center and anticenter directions 
differs between the two
studies. In addition, our sample size is only $\sim 10$\% the size of the G10 sample, and sample size-related 
biases can  also affect this comparison accordingly.

 Otherwise,  those stars with both determinations $E(B-V)$ and $E(J-K)$, only $27$~\% present consistent values ($E(B-V)\sim0.5~E(J-K)$), which
can be produced by the errors in the photometry (see Sec.~\ref{method}), and/or errors in determination of \teff and their respective errors . Therefore, it is important to take these values with
caution.


%


\subsection{The iron and Li  abundances and three Li-rich giants}\label{CAP:FE}

\begin{figure}[!h]
   \centering
   \includegraphics[scale=0.8]{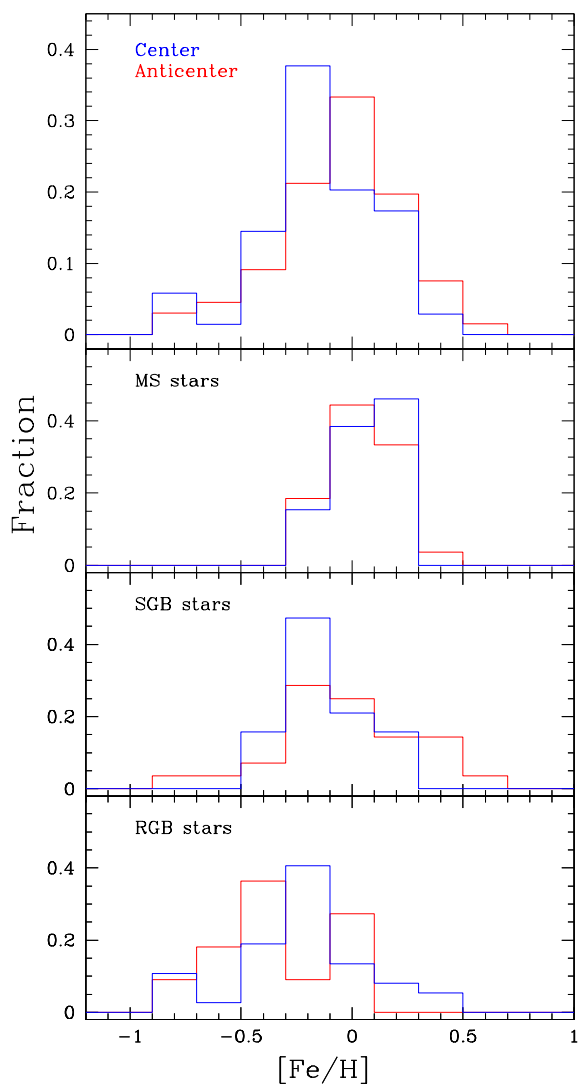}
   \caption{\emph{Upper panel}: histogram of the iron abundances \fe in the CoRoT fields. \emph{Lower panels}: 
   histograms of the iron abundances \fe    for the different evolutionary stages. A relation between 
   \fe and evolutionary status is clearly present for the stars in our sample.  Stars with
   abnormal lithium behavior are presented using filled circles.}
              \label{fig:histoFe}%
    \end{figure}

Figure~\ref{fig:histoFe} shows that our values of  \fe strongly agree
with typical values found in the Galactic disk, which range from
$-1.3$ to $+0.4$ \citep{Reddy_etal2006,Bensby_etal2007,Melendez_etal2008}. In fact, as was mentioned before,
differences are expected  in the distributions of \fe  between both CoRoT fields. As shown in Fig.~2,
the  \cen  ~field is composed of stars  with lower \fe values
than the \anti ~field. Stars with \fe$<0.0$ represent  $75\%$ and $56\%$ of the sample in the \cen
~and \anti ~fields, respectively.
Following this point, stars in those CoRoT fields also present important differences 
in the distribution of \teff, {\bf \logg}, and evolutionary stages, suggesting a link with \fe.
This can be seen when the histograms for each evolutionary stage are analyzed (see Fig.~\ref{fig:histoFe}).
The mean values of \fe for stars in the MS, SGB, and RGB of the \cen ~field are $+0.05$, $-0.16$, and
$-0.26$, respectively. For the \anti ~field, the corresponding mean values are $+0.04$, $-0.02$, and $-0.41$.

Our \fe distributions show a similar behavior to that presented by \citet{Gazzano_etal2010} in an
extensive spectroscopic survey of the CoRoT field\footnote{We present measurements of \fe, whereas  G10 present the global metallicity [M/H]. The   comparison should be done with caution since both quantities do not represent
the same abundance. }. Similarly, our results show that most
stars on the MS present  solar \fe values, whereas most stars in evolved stages have
low metallicities. The relation  between \fe,  evolutionary stages, and the
two different Galactic directions observed by CoRoT
found in \citet{Gazzano_etal2010} , which we confirmed, could be explained by the
metallicity gradient found in the Galactic disk
\citep{Pedicelli_etal2009,Friel_etal2010,Luck_etal2011}.\footnote{Our
stellar sample comprises stars in early and
evolved stages and they present a  narrow interval in apparent magnitudes $V$, which implies a
spread in absolute magnitudes and distances.}

\begin{figure}[!h]
   \centering
   \includegraphics[scale=0.8]{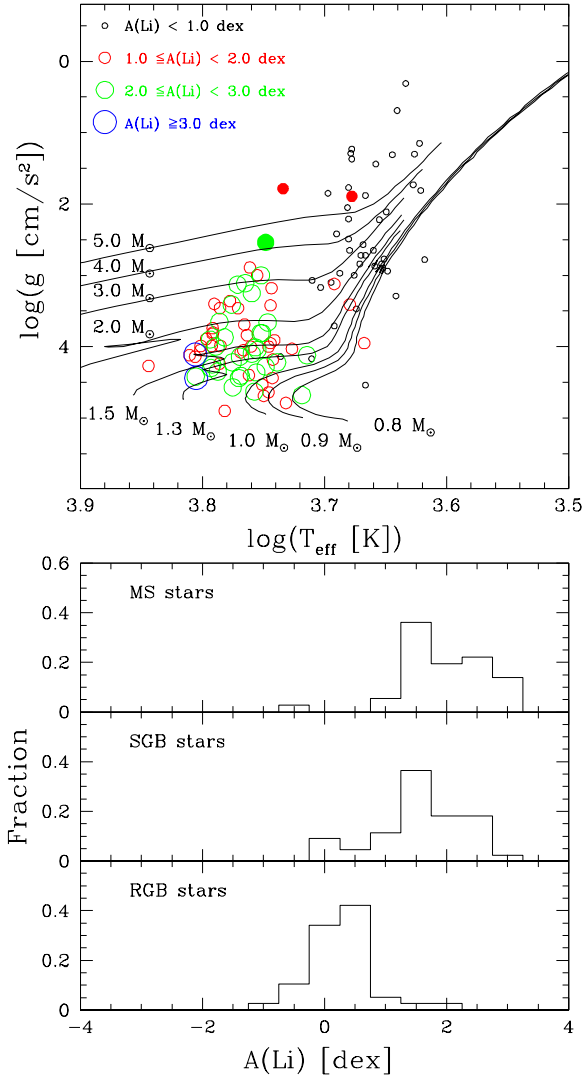}
  \caption{\emph{Upper panel}: distribution of \li along the HR diagram for the
stars of the present sample. Symbol size is proportional to the value of \li. The stars
showing an abnormal lithium behavior are presented using filled circles. Evolutionary tracks are 
the same as in Fig.~\ref{fig:HR}.  \emph{Lower panels}:  histograms of \li for  different evolutionary stages.}
             \label{fig:HR_ALI}%
    \end{figure}

 The behavior of the lithium abundances for the stars in our sample is
shown in Fig.~\ref{fig:HR_ALI}, with the \li distribution along the HR
diagram in the upper panel and histograms for different
evolutionary stages in the lower panel. The
\li measurements for the present stellar sample clearly follows the
well-established scenario for the lithium behavior at the referred
evolutionary phases \citep
{Luck1977,Boesgaard_Tripicco1986,Soderblom_etal1993,Wallerstein_etal1994,deMedeiros_etal1997,Lebre_etal1999,deMedeiros_etal2000,Melendez_etal2010}.

Indeed, the stellar lithium content is extremely sensitive to the
physical conditions inside stars.
As well established, the surface Li abundance is further depleted after stars leave the MS and undergo the first dredge-up
\citep{Iben1967a,Iben1967b}.
As a result, RGB stars essentially exhibit
 low \li \citep{Brown_etal1989}.
Nevertheless, an increasing list of studies report the discovery of
giant stars violating this rule \citep[e.g.,][]{Wallerstein_Sneden1982,
Brown_etal1989,Pilachowski_etal2000,Martell_Shetrone2013},
the so-called lithium-rich giant stars, which present atypically large
lithium abundances, in contrast to theoretical predictions.
 Three stars in the present CoRoT sample show this an abnormal lithium
behavior: CoRoT ID $100537408$, with an \li of $1.45\pm0.46$; CoRoT ID $101358013$, with an \li of $2.27\pm0.21$; and CoRoT
ID $101555541$, with an \li of $1.13\pm0.18$ (see Fig.~\ref{fig:HR_ALI}).

\begin{figure}[!h]
   \centering
   \includegraphics[scale=.8]{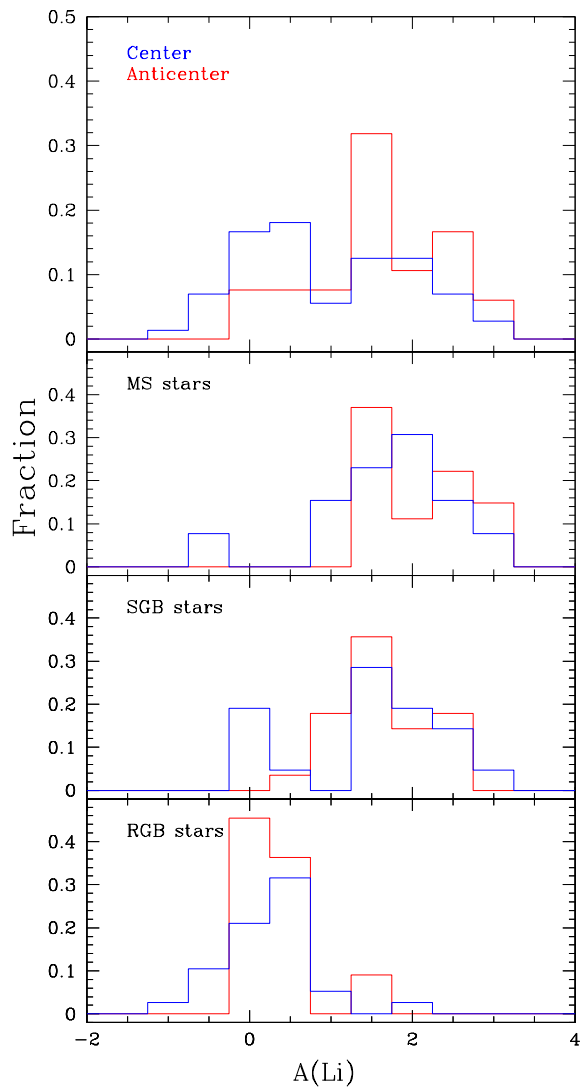}
  \caption{ \emph{Upper panel}: histogram of the lithium abundances \li in the CoRoT fields with stars 
  segregated in the two Galactic directions, with blue indicating the Center and red the Anti--center. 
   \emph{Lower panels}: histograms of A(Li) for  different evolutionary stages. }
             \label{fig:histoALI}%
    \end{figure}

 Figure~\ref{fig:histoALI} shows the distributions of \li for the whole sample
located in the Galactic center and anti--center directions. The upper panel of Fig.~\ref{fig:histoALI} provides 
an indication that a difference may be present, with the highest lithium content in the anti-center direction
and an apparent bimodal distribution in the center direction. However, the
histograms for the stars segregated by MS, SGB, and RGB evolutionary stages (bottom panel) show no statistically clear
difference between Galactic center and anti--center. Indeed, the
distributions of \li  for stars separated by evolutionary stages
seem  to indicate that the behavior of the lithium content
observed along the HR diagram follows the same trend irrespective of the
location of stars in Galactic center or anti-center directions.

%



%



\section{Conclusions}

We present physical parameters (\teff, \logg, \vmic, and \vsini), dynamical 
properties (\vrad), and chemical abundances ( \fe and \li) for a sample of  138 stars stars located in the 
CoRoT Exoplanetary fields LRc01 and LRa01, based on observations collected with UVES/VLT at ESO
and Hydra/Blanco at CTIO. The derived parameters allow us to  characterize 
physically the observed stellar sample, with the stars located in different regions of the HR diagram, from the 
MS to well-evolved stages, including the SGB and RGB. We also provide estimates of possible 
errors in the temperatures derived using photometric calibrations, and also of reddening  values 
for the stars in the aforementioned CoRoT fields.

Our results show a relation between \teff, evolutionary stage,  and \fe in the 
CoRoT fields, which is related to the color distributions in these fields. These
results are in agreement with independent photometric and spectroscopic surveys of the 
CoRoT fields.  These results give support to our spectroscopically-determined parameters.
Our chemical analysis shows that the stars in the CoRoT fields present the same 
patterns found and reported on for the Galactic disk, showing a mixture of different populations
associated with the Milky Way.

The stellar sample presents the same rotational behavior described in the literature for
different evolutionary stages and colors. Also, we provide  a calibration to derive 
\vsini from Hydra observations using the robust CCF technique (see section~\ref{CAP:VEL} and eq.~\ref{fig:CCF}). 

Finally, the present data set also represents  an important piece of work to be used as standard sample 
calibration for different programs in the context of the CoRoT mission, since, 
 among the brightest stars that comprise the CoRoT  exoplanet field targets, dozens are included 
in the list of stars  analyzed here.  It is important to remark this work also can help to increase the scientific return of other spacial missions, such as Gaia or TESS.

\begin{acknowledgements}
Research activities of the Stellar Board of the Federal University of Rio Grande do Norte are supported by 
continuous grants of CNPq and FAPERN Brazilian agencies. We also acknowledges financial support of the INCT INEspa\c{c}o. 
ICL and CEFL acknowledges postdoctoral fellowship of the CNPq; CC, SCM, CEFL, SV and GPO acknowledge
graduate fellowships of the CAPES agency. This work was partially supported by the German Deutsche
Forschungsgemeinschaft, DFG project number Ts 17/201.
Support for CC and MC is provided by the Chilean Ministry for the Economy, Development,
and Tourism's Programa Iniciativa Cient\'ifica Milenio through grant IC  120009, awarded to the Millennium Institute of Astrophysics (MAS); by Proyecto Basal PFB-
06/2007; and by Proyecto FONDECYT Regular  \#1141141.  This research has made use of the NASA/IPAC Extragalactic Database (NED) which is operated by the Jet Propulsion Laboratory, California Institute of Technology, under contract with the National Aeronautics and Space Administration.
\end{acknowledgements}

\bibliographystyle{aa}
\bibliography{ccortesabib.bib.bib}
\begin{landscape}
\tiny
\begin{center}
\begin{longtable}{lrrcccccccccc}
\caption{CoRoT IDs, ephemerides, details
of  observations, photometry, and estimated luminosity classes (LC) for the \uvesstars.} \label{TAB:FUVES} \\

\hline \hline \\[-2ex]
\multicolumn{1}{l}{CoRoT ID} &
\multicolumn{1}{c}{RA} &
\multicolumn{1}{c}{DEC} &
\multicolumn{1}{c}{Heliocentric Julian Date} &
\multicolumn{1}{c}{$\rm{T_{exp}}$} &
\multicolumn{1}{c}{$S/N$} &
\multicolumn{1}{c}{$V$} &
\multicolumn{1}{c}{$(B-V)$} &
\multicolumn{1}{c}{$J$} &
\multicolumn{1}{c}{$H$} &
\multicolumn{1}{c}{$K$} &
\multicolumn{1}{c}{LC} &
\multicolumn{1}{c}{Field} \\[0.5ex] 
     & (J2000) & (J2000) & (HJD)&       (s)     &               &       (mag)   &       (mag)   &       (mag)   & (mag)   &       (mag)   &        &      \\


        \hline
   \\[-1.8ex]
\endfirsthead

\multicolumn{3}{c}{{\tablename} \thetable{} -- Continued} \\[0.5ex]
  \hline \hline \\[-2ex]
\multicolumn{1}{l}{CoRoT ID} &
\multicolumn{1}{c}{RA} &
\multicolumn{1}{c}{DEC} &
\multicolumn{1}{c}{Heliocentric Julian Date} &
\multicolumn{1}{c}{$\rm{T_{exp}}$} &
\multicolumn{1}{c}{$S/N$} &
\multicolumn{1}{c}{$V$} &
\multicolumn{1}{c}{$(B-V)$} &
\multicolumn{1}{c}{$J$} &
\multicolumn{1}{c}{$H$} &
\multicolumn{1}{c}{$K$} &
\multicolumn{1}{c}{LC} &
\multicolumn{1}{c}{Field} \\[0.5ex] 
     &(J2000) &(J2000)  & (HJD)&        (s)     &               &       (mag)   &       (mag)   &       (mag)   & (mag)   &       (mag)   &        &      \\

        \hline
  \\[-1.8ex]
\endhead

  \multicolumn{3}{l}{{Continue in the next page\ldots}} \\
\endfoot

\hline \hline 
  \\[-1.8ex] 
\endlastfoot

100532655       &$      19:23:03.\!118  $&$     +01:34:55.\!60  $&      2006    Apr     23      &$      1000    $&$     128     $&$     12.\!03 $&$     0.\!93  $&$     10.\!38 $&$     10.\!01 $&$     9.\!93  $&      IV      &       LRc01   \\
100537408       &$      19:23:04.\!841  $&$     +01:46:24.\!49  $&      2006    May     16      &$      1000    $&$     130     $&$     12.\!23 $&$     0.\!85  $&$     10.\!60 $&$     10.\!36 $&$     10.\!22 $&      III     &       LRc01   \\
100931549       &$      19:25:21.\!353  $&$     +00:11:07.\!94  $&      2006    May     23      &$      1000    $&$     129     $&$     12.\!48 $&$     0.\!89  $&$     10.\!93 $&$     10.\!58 $&$     10.\!48 $&      IV      &       LRc01   \\
100932329       &$      19:25:21.\!593  $&$     +00:11:31.\!02  $&      2006    May     23      &$      1000    $&$     200     $&$     12.\!12 $&$     1.\!01  $&$     10.\!33 $&$     9.\!88  $&$     9.\!76  $&      IV      &       LRc01   \\
101041358       &$      19:26:00.\!737  $&$     +00:36:33.\!01  $&      2006    Jun     17      &$      1000    $&$     131     $&$     12.\!43 $&$     0.\!90  $&$     10.\!71 $&$     10.\!27 $&$     10.\!18 $&      V       &       LRc01   \\
101043867       &$      19:26:01.\!891  $&$     -00:02:49.\!67  $&      2006    Jun     17      &$      1000    $&$     111     $&$     12.\!42 $&$     0.\!85  $&$     10.\!83 $&$     10.\!49 $&$     10.\!43 $&      IV      &       LRc01   \\
101056542       &$      19:26:07.\!651  $&$     +00:11:03.\!37  $&      2006    Jun     17      &$      1000    $&$     122     $&$     12.\!43 $&$     0.\!67  $&$     11.\!14 $&$     10.\!90 $&$     10.\!85 $&      IV      &       LRc01   \\
101076647       &$      19:26:17.\!654  $&$     +00:12:09.\!00  $&      2006    Apr     19      &$      1000    $&$     102     $&$     12.\!43 $&$     0.\!90  $&$     10.\!73 $&$     10.\!40 $&$     10.\!32 $&      IV      &       LRc01   \\
101102758       &$      19:26:27.\!502  $&$     +00:32:20.\!15  $&      2006    Apr     23      &$      1000    $&$     124     $&$     12.\!43 $&$     1.\!00  $&$     10.\!62 $&$     10.\!24 $&$     10.\!10 $&      IV      &       LRc01   \\
101128747       &$      19:26:37.\!291  $&$     +00:02:58.\!96  $&      2006    Jun     17      &$      1000    $&$     141     $&$     12.\!00 $&$     0.\!87  $&$     10.\!46 $&$     10.\!22 $&$     10.\!11 $&      III     &       LRc01   \\
101204408       &$      19:27:05.\!707  $&$     +00:32:05.\!06  $&      2006    Jun     17      &$      1000    $&$     142     $&$     13.\!58 $&$     1.\!52  $&$     10.\!74 $&$     10.\!00 $&$     9.\!86  $&      V       &       LRc01   \\
101208246       &$      19:27:07.\!138  $&$     +00:27:19.\!33  $&      2006    Jun     17      &$      1000    $&$     109     $&$     12.\!43 $&$     1.\!02  $&$     10.\!61 $&$     10.\!20 $&$     10.\!08 $&      IV      &       LRc01   \\
101231832       &$      19:27:16.\!032  $&$     +00:17:35.\!59  $&      2006    Jun     17      &$      1000    $&$     150     $&$     12.\!13 $&$     0.\!90  $&$     10.\!47 $&$     10.\!08 $&$     10.\!02 $&      IV      &       LRc01   \\
101462309       &$      19:28:53.\!256  $&$     -00:04:41.\!20  $&      2006    Jun     18      &$      1000    $&$     130     $&$     12.\!13 $&$     0.\!89  $&$     10.\!55 $&$     10.\!22 $&$     10.\!15 $&      V       &       LRc01   \\
101476063       &$      19:28:58.\!836  $&$     +00:40:16.\!72  $&      2006    Jun     18      &$      1000    $&$     105     $&$     12.\!20 $&$     0.\!97  $&$     10.\!46 $&$     10.\!07 $&$     9.\!93  $&      III     &       LRc01   \\
101478005       &$      19:28:59.\!618  $&$     -00:07:21.\!76  $&      2006    Jun     18      &$      1000    $&$     120     $&$     12.\!24 $&$     0.\!99  $&$     10.\!52 $&$     10.\!07 $&$     9.\!99  $&      IV      &       LRc01   \\
101565378       &$      19:29:35.\!662  $&$     -00:12:26.\!06  $&      2006    May     30      &$      1000    $&$     174     $&$     12.\!33 $&$     0.\!80  $&$     10.\!87 $&$     10.\!56 $&$     10.\!47 $&      IV      &       LRc01   \\
101613938       &$      19:29:56.\!532  $&$     +00:01:31.\!04  $&      2006    May     30      &$      1000    $&$     167     $&$     12.\!23 $&$     0.\!86  $&$     10.\!70 $&$     10.\!40 $&$     10.\!35 $&      V       &       LRc01   \\
101697676       &$      19:30:35.\!563  $&$     +00:03:20.\!05  $&      2006    May     30      &$      1000    $&$     153     $&$     12.\!37 $&$     0.\!80  $&$     10.\!99 $&$     10.\!67 $&$     10.\!60 $&      V       &       LRc01   \\
102585563       &$      06:41:02.\!086  $&$     -00:33:16.\!38  $&      2006    Sept    07      &$      1000    $&$     166     $&$     16.\!12 $&$     1.\!08  $&$     13.\!71 $&$     13.\!09 $&$     12.\!99 $&      V       &       LRa01   \\
102585613       &$      06:41:02.\!184  $&$     +00:19:38.\!39  $&      2006    Sept    06      &$      1000    $&$     188     $&$     12.\!34 $&$     0.\!44  $&$     11.\!31 $&$     11.\!11 $&$     11.\!03 $&      III     &       LRa01   \\
102589564       &$      06:41:08.\!995  $&$     +00:48:20.\!41  $&      2006    Sept    07      &$      1000    $&$     133     $&$     12.\!46 $&$     0.\!50  $&$     11.\!49 $&$     11.\!29 $&$     11.\!21 $&      V       &       LRa01   \\
102591896       &$      06:41:12.\!994  $&$     -00:45:13.\!54  $&      2006    Sept    03      &$      1000    $&$     150     $&$     12.\!25 $&$     0.\!54  $&$     11.\!28 $&$     11.\!09 $&$     11.\!00 $&      V       &       LRa01   \\
102603174       &$      06:41:31.\!006  $&$     -00:57:44.\!14  $&      2006    Sept    10      &$      1000    $&$     138     $&$     12.\!08 $&$     0.\!56  $&$     11.\!14 $&$     10.\!86 $&$     10.\!81 $&      V       &       LRa01   \\
102604055       &$      06:41:32.\!222  $&$     +00:59:10.\!75  $&      2006    Sept    10      &$      1000    $&$     126     $&$     12.\!35 $&$     0.\!46  $&$     11.\!41 $&$     11.\!21 $&$     11.\!15 $&      V       &       LRa01   \\
102605405       &$      06:41:34.\!037  $&$     +00:00:30.\!24  $&      2006    Sept    13      &$      1000    $&$     211     $&$     12.\!08 $&$     0.\!60  $&$     10.\!97 $&$     10.\!69 $&$     10.\!63 $&      V       &       LRa01   \\
102606185       &$      06:41:35.\!021  $&$     +00:54:44.\!68  $&      2006    Sept    13      &$      1000    $&$     105     $&$     13.\!44 $&$     0.\!56  $&$     12.\!39 $&$     12.\!11 $&$     12.\!10 $&      V       &       LRa01   \\
102611980       &$      06:41:42.\!962  $&$     -00:24:33.\!84  $&      2006    Sept    13      &$      1000    $&$     117     $&$     12.\!41 $&$     0.\!65  $&$     11.\!30 $&$     11.\!00 $&$     10.\!97 $&      V       &       LRa01   \\
102614844       &$      06:41:46.\!882  $&$     +00:21:40.\!72  $&      2006    Sept    22      &$      1000    $&$     91      $&$     12.\!50 $&$     0.\!83  $&$     11.\!13 $&$     10.\!78 $&$     10.\!72 $&      IV      &       LRa01   \\
102616719       &$      06:41:49.\!716  $&$     -01:32:02.\!26  $&      2006    Aug     31      &$      1000    $&$     123     $&$     12.\!31 $&$     0.\!59  $&$     11.\!30 $&$     11.\!08 $&$     10.\!99 $&      V       &       LRa01   \\
102618948       &$      06:41:52.\!913  $&$     +00:24:13.\!86  $&      2006    Sept    22      &$      1000    $&$     134     $&$     12.\!01 $&$     0.\!66  $&$     10.\!88 $&$     10.\!63 $&$     10.\!55 $&      III     &       LRa01   \\
102620828       &$      06:41:55.\!704  $&$     -00:59:37.\!90  $&      2006    Sept    24      &$      1000    $&$     120     $&$     12.\!36 $&$     0.\!85  $&$     10.\!90 $&$     10.\!47 $&$     10.\!37 $&      III     &       LRa01   \\
102654716       &$      06:42:42.\!254  $&$     +00:47:29.\!00  $&      2006    Sept    23      &$      1000    $&$     171     $&$     12.\!15 $&$     0.\!55  $&$     11.\!06 $&$     10.\!83 $&$     10.\!74 $&      III     &       LRa01   \\
102657182       &$      06:42:45.\!377  $&$     -00:32:21.\!05  $&      2006    Sept    24      &$      1000    $&$     116     $&$     12.\!35 $&$     0.\!78  $&$     10.\!98 $&$     10.\!60 $&$     10.\!55 $&      III     &       LRa01   \\
102658181       &$      06:42:46.\!610  $&$     +00:53:40.\!09  $&      2006    Sept    24      &$      1000    $&$     103     $&$     12.\!38 $&$     0.\!71  $&$     11.\!12 $&$     10.\!80 $&$     10.\!73 $&      V       &       LRa01   \\
102659670       &$      06:42:48.\!432  $&$     +00:29:03.\!59  $&      2006    Sept    25      &$      1000    $&$     129     $&$     12.\!01 $&$     0.\!51  $&$     11.\!07 $&$     10.\!85 $&$     10.\!81 $&      III     &       LRa01   \\
102663892       &$      06:42:53.\!772  $&$     -01:17:53.\!56  $&      2006    Sept    27      &$      1000    $&$     145     $&$     12.\!36 $&$     0.\!60  $&$     11.\!31 $&$     11.\!03 $&$     10.\!98 $&      V       &       LRa01   \\
102669038       &$      06:43:00.\!288  $&$     +00:49:01.\!52  $&      2006    Sept    26      &$      750     $&$     112     $&$     12.\!38 $&$     0.\!46  $&$     11.\!37 $&$     11.\!17 $&$     11.\!09 $&      III     &       LRa01   \\
102669801       &$      06:43:01.\!198  $&$     +00:14:26.\!81  $&      2006    Sept    25      &$      1000    $&$     111     $&$     12.\!34 $&$     0.\!66  $&$     11.\!10 $&$     10.\!80 $&$     10.\!73 $&      III     &       LRa01   \\
102676872       &$      06:43:09.\!950  $&$     -01:27:26.\!03  $&      2006    Sept    25      &$      1000    $&$     120     $&$     12.\!19 $&$     0.\!98  $&$     10.\!28 $&$     9.\!81  $&$     9.\!68  $&      III     &       LRa01   \\
102678564       &$      06:43:12.\!110  $&$     -00:26:59.\!78  $&      2006    Sept    25      &$      1000    $&$     130     $&$     12.\!11 $&$     0.\!51  $&$     11.\!13 $&$     10.\!93 $&$     10.\!86 $&      V       &       LRa01   \\
102679796       &$      06:43:13.\!630  $&$     +00:45:08.\!89  $&      2006    Sept    25      &$      1000    $&$     127     $&$     12.\!04 $&$     0.\!50  $&$     11.\!06 $&$     10.\!83 $&$     10.\!79 $&      IV      &       LRa01   \\
102686019       &$      06:43:21.\!432  $&$     -00:04:27.\!05  $&      2006    Sept    25      &$      1000    $&$     130     $&$     12.\!20 $&$     0.\!50  $&$     11.\!30 $&$     11.\!07 $&$     11.\!02 $&      V       &       LRa01   \\
102687759       &$      06:43:23.\!638  $&$     +00:12:21.\!89  $&      2006    Sept    26      &$      1000    $&$     140     $&$     12.\!21 $&$     0.\!56  $&$     11.\!14 $&$     10.\!86 $&$     10.\!85 $&      V       &       LRa01   \\
102692093       &$      06:43:29.\!011  $&$     +01:02:31.\!52  $&      2006    Oct     01      &$      1000    $&$     132     $&$     12.\!47 $&$     0.\!61  $&$     11.\!25 $&$     10.\!94 $&$     10.\!87 $&      III     &       LRa01   \\
102705308       &$      06:43:44.\!971  $&$     +00:55:50.\!41  $&      2006    Sept    27      &$      1000    $&$     167     $&$     12.\!39 $&$     0.\!96  $&$     10.\!59 $&$     10.\!09 $&$     10.\!00 $&      III     &       LRa01   \\
102709247       &$      06:43:50.\!210  $&$     -01:11:03.\!01  $&      2006    Sept    27      &$      1000    $&$     142     $&$     12.\!46 $&$     0.\!77  $&$     11.\!25 $&$     10.\!91 $&$     10.\!88 $&      III     &       LRa01   \\
102718064       &$      06:44:02.\!532  $&$     +00:27:04.\!14  $&      2006    Sept    27      &$      1000    $&$     167     $&$     15.\!91 $&$     0.\!73  $&$     14.\!41 $&$     14.\!26 $&$     13.\!91 $&      V       &       LRa01   \\
102738854       &$      06:44:31.\!608  $&$     -00:18:21.\!10  $&      2006    Sept    27      &$      1000    $&$     126     $&$     12.\!25 $&$     0.\!63  $&$     11.\!12 $&$     10.\!85 $&$     10.\!80 $&      IV      &       LRa01   \\
102740520       &$      06:44:33.\!814  $&$     -00:50:35.\!77  $&      2006    Sept    05      &$      1000    $&$     103     $&$     12.\!40 $&$     0.\!79  $&$     11.\!19 $&$     10.\!88 $&$     10.\!82 $&      V       &       IRa01   \\
102741215       &$      06:44:34.\!738  $&$     -01:55:45.\!69  $&      2006    Sept    05      &$      1000    $&$     126     $&$     12.\!41 $&$     0.\!63  $&$     11.\!25 $&$     10.\!97 $&$     10.\!91 $&      V       &       LRa01   \\
102764866       &$      06:45:06.\!610  $&$     -01:05:31.\!81  $&      2006    Sept    28      &$      1000    $&$     150     $&$     12.\!14 $&$     0.\!63  $&$     11.\!01 $&$     10.\!71 $&$     10.\!63 $&      IV      &       LRa01   \\
102772347       &$      06:45:16.\!702  $&$     -00:56:54.\!06  $&      2006    Sept    28      &$      1000    $&$     142     $&$     12.\!14 $&$     0.\!66  $&$     10.\!97 $&$     10.\!65 $&$     10.\!59 $&      IV      &       LRa01   \\
                                                                                                                                                                                                                                        
\end{longtable}
\end{center}

\normalsize

\end{landscape}


\begin{table}[h]

\caption{Setup used for Hydra observations.}
\centering
\begin{tabular}{rcccc}
\hline \hline
Field   &       Number of exposures     &       $T_{\rm{exp}}$  &       Heliocentric Julian Date     &       Filter  \\
            &           &       (s)     &       (HDJ)   &               \\[0.5ex]
\hline
$LRa01$ &       $2$     &       $2700$  &       2009 Oct 04     &       $E5187$ \\
$LRc01$     &   $2$     &       $2700$  &       2009 Oct 04     &       $E5187$ \\
$LRa01$ &       $2$     &       $2200$  &       2009 Oct 04     &       $E6757$ \\
$LRc01$     &   $2$     &       $2850$  &       2009 Oct 04     &       $E6757$ \\

\hline                  
\end{tabular}
\label{TAB:OBSHYDRA}
\end{table}

\begin{landscape}
\tiny
\begin{center}
\begin{longtable}{lrrcccccccc}
\caption{ The main characteristics for the \hydrastars.} \label{TAB:FHYDRA} \\

\hline \hline \\[-2ex]
\multicolumn{1}{l}{CoRoT ID} &
\multicolumn{1}{c}{RA} &
\multicolumn{1}{c}{DEC} &
\multicolumn{1}{c}{$S/N$} &
\multicolumn{1}{c}{$V$} &
\multicolumn{1}{c}{$(B-V)$} &
\multicolumn{1}{c}{$J$} &
\multicolumn{1}{c}{$H$} &
\multicolumn{1}{c}{$K$} &
\multicolumn{1}{c}{LC} &
\multicolumn{1}{c}{Field}
\\[0.5ex] 
     &(J2000) &(J2000)  & &     (mag)   &       (mag)   &       (mag)   & (mag)   &       (mag)   &        &      \\


        \hline
   \\[-1.8ex]
\endfirsthead

\multicolumn{3}{c}{{\tablename} \thetable{} -- Continued} \\[0.5ex]
\hline \hline \\[-2ex]
\multicolumn{1}{l}{CoRoT ID} &
\multicolumn{1}{c}{RA} &
\multicolumn{1}{c}{DEC} &
\multicolumn{1}{c}{$S/N$} &
\multicolumn{1}{c}{$V$} &
\multicolumn{1}{c}{$(B-V)$} &
\multicolumn{1}{c}{$J$} &
\multicolumn{1}{c}{$H$} &
\multicolumn{1}{c}{$K$} &
\multicolumn{1}{c}{LC} &
\multicolumn{1}{c}{Field}
\\[0.5ex] 
     &(J2000) &(J2000) &   &    (mag)   &       (mag)   &       (mag)   & (mag)   &       (mag)   &        &      \\

        \hline
  \\[-1.8ex]
\endhead

  \multicolumn{3}{l}{{Continue in the next page\ldots}} \\
\endfoot

\hline \hline 
  \\[-1.8ex] 
\endlastfoot
101057962       &$      19:26:08.33                                     $&$     -00:25:22.80                                    $&$     167     $&$     12.69   $&$     1.44    $&$     9.95    $&$     9.22    $&$     9.07    $&      V       &       LRc01   \\
101066444       &$      19:26:12.18                                     $&$     +00:01:28.60                                    $&$     123     $&$     13.37   $&$     1.81    $&$     10.04   $&$     9.25    $&$     9.03    $&      II      &       LRc01   \\
101124344       &$      19:26:35.61                                     $&$     -00:27:25.00                                    $&$     187     $&$     13.18   $&$     1.24    $&$     10.92   $&$     10.39   $&$     10.26   $&      III     &       LRc01   \\
101154362       &$      19:26:46.93                                     $&$     -00:12:23.80                                    $&$     170     $&$     13.13   $&$     1.34    $&$     10.77   $&$     10.13   $&$     9.98    $&      V       &       LRc01   \\
101165983       &$      19:26:51.33                                     $&$     -00:11:36.20                                    $&$     85      $&$     13.55   $&$     1.40    $&$     10.98   $&$     10.28   $&$     10.12   $&      V       &       LRc01   \\
101178248       &$      19:26:55.94                                     $&$     +00:06:04.30                                    $&$     168     $&$     12.38   $&$     1.48    $&$     9.71    $&$     9.04    $&$     8.83    $&      V       &       LRc01   \\
\end{longtable}
\end{center}

\normalsize

\end{landscape}


\begin{table}[h]
\centering
\caption{ The main characteristics  and rotation velocities \vsini of calibrator stars.}

\begin{tabular}{lcrccrrrc}
\hline \hline
ID      &       \# spectra      &       $T_{\rm{exp}}$  &       Date    &       Hydra Filter & $V$ & $(B-V)$ &\vsini& References\\
            &           &       (s)     &               &  &(mag) &(mag) &\kms   &       \\[0.5ex]
\hline
HD~96843          &     $2$     &       $600$   &       2009 Oct 04     &       $E5187$ & $8.36$ & $0.66$ & $13.0$& $1$ \\
HD~101117  &    $2$     &       $600$   &       2009 Oct 04     &       $E5187$ & $7.87$ & $0.59$ & $9.0$ & $2$ \\
HD~113226       &       $8$     &       $7$       &     2009 Oct 04     &       $E5187$ & $2.83$ & $0.93$ & $3.0$ &$3$  \\
HD~113553  &    $3$     &       $600$   &       2009 Oct 04     &       $E5187$ & $8.33$ & $0.68$ & $12.0$ & $1$\\
HD~141710       &       $2$     &       $600$   &       2009 Oct 04     &       $E5187$ & $8.93$ & $0.14$ & $30.0$& $1$ \\
HD~150108       &       $2$     &       $600$   &       2009 Oct 04     &       $E5187$ & $8.23$ & $0.84$ & $21.0$ & $1$\\
HD~165760  &    $3$     &       $30$    &       2009 Oct 04     &       $E5187$ & $4.67$ & $0.95$ & $2.6$  &$3$\\
\hline
\multicolumn{9}{l}{{\bf References:} (1) \citealt{Cutispoto_etal2002}; (2) \citealt{Nordstrom_etal2004}; (3) \citealt{Melo_etal2001}}\\
\end{tabular}
 
\label{TAB:OBSCALIB}
\end{table}


%
 \begin{table}[h]
 \caption{Observed broadening  $\sigma_{obs}$  and  $\sigma_t$ for template stars.}
 \centering
 \begin{tabular}{lcccc}
 \hline \hline
 ID     & \# spectra    &       $<\sigma_{obs}^{P}>$    &       $\Delta\sigma_{obs}$    &       $<\sigma_t>$    \\
            &           &       \kms    &       \kms    &       \kms    \\[0.5ex]
 \hline
 HD~113226      &       $8$     &       $29.9$  &       $0.6$   &       $21.2$  \\
 HD~165760      &       $3$     &       $30.5$  &       $0.1$   &       $21.5$         \\

 \hline                 
 \end{tabular}
 \label{tab:Template}
 \end{table}
%
 \begin{table}[h]
 \caption{Observed broadening $\sigma_{obs}$ and $\sigma_{obs0}$ for calibrator stars.}
\begin{tabular}{lccccccc}
 \hline \hline%
  & \multicolumn{4}{c}{Template}&       \multicolumn{2}{c}{} & \\                                                       
& \multicolumn{2}{c}{$\rm{HD~113226}^{a}$}      &\multicolumn{2}{c}{$\rm{HD~165760}^{b}$}       &               &                       &               \\
  [0.5ex]
 \hline
Calibrator & $<\sigma_{obs}>$   &$\Delta\sigma_{obs}$   &       $<\sigma_{obs}>$         &$\Delta\sigma_{obs}$   &               $<\sigma_{obs0}>^a$     &       $<\sigma_{obs0}>^b$             &       $<\sigma_{obs0}>$\\
        &       \kms    &       \kms    &               \kms    &       \kms    &       \kms    &       \kms    &       \kms    \\[0.5ex]
 
 \hline
 HD~96843       &$      33.4    $&$     0.3     $&$     33.5    $&$     0.1     $&$     25.8    $&$     25.7    $&$     25.8    $\\
 HD~101117      &$      32.6    $&$     0.4     $&$     32.8    $&$     0.1     $&$     24.7    $&$     24.7    $&$     24.7 $       \\
 HD~113226      &$      ...     $&$     ...     $&$     30.2    $&$     0.5     $&$     ...     $&$     21.2    $&$     21.2    $\\
 HD~113553      &$      33.5    $&$     0.4     $&$     33.7    $&$     0.1     $&$     25.9    $&$     25.9    $&$     25.9    $\\
 HD~141710      &$      46.2    $&$     0.3     $&$     46.2    $&$     0.2     $&$     41.1    $&$     40.9    $&$     41.0    $\\
 HD~150108      &$      37.5    $&$     0.4     $&$     37.7    $&$     0.3     $&$     31.0    $&$     30.9    $&$     31.0    $\\
 HD~165760      &$      30.3    $&$     0.5     $&$     ...     $&$     ...     $&$     21.7    $&$     ...     $&$     21.7    $\\

 \hline                 
 \end{tabular}

 \label{tab:Calibrators}
 \end{table}
%
%

\begin{landscape}
        \tiny
        \begin{center}
                \begin{longtable}{crrrrrrrrrrrrcc}
                        \caption{ Stellar parameters for the stars in our sample. Refer to sections \ref{method} and \ref{CAP:RESULTS} for details} \label{TAB:ALLPARAMETERS} \\
                        
                        \hline \hline \\[-2ex]
                        \multicolumn{1}{l}{CoRoT ID} &
                        \multicolumn{1}{c}{$T_{\rm{eff}}(B-V)$} &
                        \multicolumn{1}{c}{$T_{\rm{eff}}(J-K)$} &
                        \multicolumn{1}{c}{\teff} &
                        \multicolumn{1}{c}{\logg} &
                        \multicolumn{1}{c}{\vmic} &
                        \multicolumn{1}{c}{\fe} &
                        \multicolumn{1}{c}{Fe lines} &
                        \multicolumn{1}{c}{\li} &
                        \multicolumn{1}{c}{\vsini} &
                        \multicolumn{1}{c}{\vrad} &
                        \multicolumn{1}{c}{$E(B-V)$} &
                        \multicolumn{1}{c}{$E(J-K)$} &
                        \multicolumn{1}{c}{Evol. Status}&
                        \multicolumn{1}{c}{Inst}
                        \\[0.5ex] 
                        & (K)& (K) &(K) &       (dex)   & (km/s)                & (dex)           &FeI,FeII & (dex) &     \kms    &\kms   & (mag) &       (mag)& &               \\                      
                        
                        \hline
                        \\[-1.8ex]
                        \endfirsthead
                        
                        \multicolumn{3}{c}{{\tablename} \thetable{} -- Continued} \\[0.5ex]
                        \hline \hline \\[-2ex]
                        \multicolumn{1}{l}{CoRoT ID} &
                        \multicolumn{1}{c}{$T_{\rm{eff}}(B-V)$} &
                        \multicolumn{1}{c}{$T_{\rm{eff}}(J-K)$} &
                        \multicolumn{1}{c}{\teff} &
                        \multicolumn{1}{c}{\logg} &
                        \multicolumn{1}{c}{\vmic} &
                        \multicolumn{1}{c}{\fe} &
                        \multicolumn{1}{c}{Fe lines} &
                        \multicolumn{1}{c}{\li} &
                        \multicolumn{1}{c}{\vsini} &
                        \multicolumn{1}{c}{\vrad} &
                        \multicolumn{1}{c}{$E(B-V)$} &
                        \multicolumn{1}{c}{$E(J-K)$} &
                        \multicolumn{1}{c}{Evol. Status}&
                        \multicolumn{1}{c}{Inst}
                        \\[0.5ex] 
                        & (K)& (K) &(K) &       (dex)   & (km/s)                & (dex)           &FeI,FeII & (dex) &     \kms    &\kms   & (mag) &       (mag)& &               \\                      
                        \hline
                        \\[-1.8ex]
                        \endhead
                        
                        \multicolumn{3}{l}{{Continue in the next page\ldots}} \\
                        \endfoot
                        
                        \hline \hline 
                        \\[-1.8ex] 
                        \endlastfoot

 &\multicolumn{11}{c}{Main Sequence stars}& & &\\
  & & & & & & & & & & & & & & \\
100931549       &$      5078    $&$     5571    $&$     5870    ^{+     105     }_{-    90      }$&$    4.08    ^{+     0.13    }_{-    0.13    }$&$    1.65    ^{+     0.09    }_{-    0.09    }$&$    -0.16   \pm     0.04    $&$     53,7    $&$     1.96    \pm     0.12    $&$     2       ^{+     1       }_{-    1       }$&$    -68.2   $&$     0.28    $&$     0.04    $&      MS      &       UVES    \\
101057962       &$      4082    $&$     3886    $&$     4640    ^{+     120     }_{-    110     }$&$    4.54    ^{+     0.49    }_{-    0.49    }$&$    1.68    ^{+     0.42    }_{-    0.36    }$&$    0.25    \pm     0.15    $&$     15,4    $&$     -0.30   \pm     0.16    $&$     3       ^{+     2       }_{-    2       }$&$    41.3    $&$     0.30    $&$     0.21    $&      MS      &       HYDRA   \\
\end{longtable}
\footnotetext[1]{The measurement of \li can not be obtained from the spectrum.}
\footnotetext[2]{Reddening values $E(B-V)$ can not be obtained since \teff$<T_{\rm{eff}}(B-V)$.}
\footnotetext[3]{Reddening values $E(J-K)$ can not be obtained since \teff$<T_{\rm{eff}}(J-K)$. }

\end{center}

\normalsize

\end{landscape}

\begin{table}[!h]
\centering
\caption{Mean Parameters of the CoRoT fields}
\begin{tabular}{lcc}
\hline \hline
& \cen &\anti \\

\hline
\multicolumn{3}{c}{All evolutionary stages}\\
\hline
$T_{\rm{eff}}(B-V)$ [K] &$4395 \pm595 $& $5507  \pm631 $ \\ 
$T_{\rm{eff}}(J-K)$ [K] &$4558 \pm711 $&$5562 \pm802 $ \\ 
\teff [K] &$5008 \pm677 $&$5693 \pm529 $ \\ 
\logg [dex] &$ 2.91\pm1.14 $&$ 3.68\pm0.88$\\
\fe [dex] &$-0.16 \pm0.28 $&$-0.06 \pm0.27 $\\
\vrad [\kms]&$-15.3 \pm38.3 $&$ -0.1\pm43.7 $\\
\end{tabular}
 
\label{tab:meanparam}
\end{table}


\begin{table}[h]
\centering
\caption{Common stars with \citet{Gazzano_etal2010}}
\begin{tabular}{lrrrrrr}
\hline \hline

 & \multicolumn{3}{c}{Our work} & \multicolumn{3}{c}{\citet{Gazzano_etal2010}}\\
ID      &       \teff   & \logg & \vsini& \teff & \logg & \vsini \\
            & K         &       [dex]   & \kms          & K             &       [dex]   & \kms\\[0.5ex]
\hline
$101102758      $&$     5760    $&$     4.00    $&$     2.0     $&$     5865    $&$     4.17    $&$     1.1     $\\
$101208246      $&$     5530    $&$     4.44    $&$     1.0     $&$     5600    $&$     4.72    $&$     5.4     $\\
$101538522      $&$     4180    $&$     1.81    $&$     3.0     $&$     4159    $&$     2.79    $&$     4.6     $\\
$101642292      $&$     4500    $&$     2.84    $&$     4.0     $&$     4757    $&$     3.20    $&$     4.1     $\\
$102657182      $&$     5525    $&$     4.18    $&$     2.0     $&$     5521    $&$     4.35    $&$     5.5     $\\
$102705009      $&$     5510    $&$     3.91    $&$     4.0     $&$     5858    $&$     4.35    $&$     2.7     $\\
$102730409      $&$     5720    $&$     4.63    $&$     3.0     $&$     5832    $&$     4.41    $&$     5.6     $\\
$102738854      $&$     5840    $&$     4.05    $&$     4.0     $&$     5901    $&$     4.29    $&$     5.7     $\\
$102740367      $&$     5700    $&$     4.36    $&$     12.0    $&$     \ldots  $&$     \ldots  $&$     15.7    $\\
$102740520      $&$     5640    $&$     3.81    $&$     4.0     $&$     5696    $&$     3.95    $&$     4.4     $\\
$102743929      $&$     5330    $&$     4.03    $&$     5.0     $&$     5002    $&$     3.30    $&$     6.0     $\\

\hline
\end{tabular}
\label{TAB:common}
\end{table}

\end{document}